\documentclass[twocolumn]{aastex63}
\usepackage{amssymb}
\usepackage{amsmath}
\usepackage{textcomp}
\usepackage{lipsum} 
\usepackage[super]{nth}

\defcitealias{Hacar2017}{H17}

\submitjournal{AJ}

\begin{document}

\turnoffeditone
\turnoffeditthree

\title{Velocity-Coherent Filaments in NGC 1333: Evidence for Accretion Flow?}

\correspondingauthor{Michael Chun-Yuan Chen}
\email{mcychen@uvic.ca}

\author[0000-0003-4242-973X]{Michael Chun-Yuan Chen}
\affiliation{Department of Physics and Astronomy, University of Victoria, Victoria, BC, V8P 1A1, Canada}

\author[0000-0002-9289-2450]{James Di Francesco}
\affiliation{Herzberg Astronomy and Astrophysics, National Research Council of Canada, 5071 West Saanich Road, Victoria, BC, V9E 2E7, Canada}
\affiliation{Department of Physics and Astronomy, University of Victoria, Victoria, BC, V8P 1A1, Canada}

\author[0000-0002-5204-2259]{Erik Rosolowsky}
\affiliation{Department of Physics, University of Alberta, Edmonton, AB, Canada}

\author[0000-0003-2628-0250]{Jared Keown}
\affiliation{Department of Physics and Astronomy, University of Victoria, Victoria, BC, V8P 1A1, Canada}

\author[0000-0002-3972-1978]{Jaime E. Pineda}
\affiliation{Max-Planck-Institut für extraterrestrische Physik, Giessenbachstrasse 1, 85748 Garching, Germany}

\author[0000-0001-7594-8128]{Rachel K. Friesen}
\affiliation{Department of Astronomy \& Astrophysics, University of Toronto, 50 St. George Street, Toronto, Ontario, Canada M5S 3H4}

\author[0000-0003-1481-7911]{Paola Caselli}
\affiliation{Max-Planck-Institut für extraterrestrische Physik, Giessenbachstrasse 1, 85748 Garching, Germany}

\author[0000-0001-6222-1712]{How-Huan Chen}
\affiliation{Department of Astronomy, University of Texas at Austin, Austin, TX 78712, USA}

\author[0000-0001-9732-2281]{Christopher D. Matzner}
\affiliation{Department of Astronomy \& Astrophysics, University of Toronto, 50 St. George Street, Toronto, Ontario, Canada M5S 3H4}

\author[0000-0003-1252-9916]{Stella S. Offner}
\affiliation{Department of Astronomy, University of Texas at Austin, Austin, TX 78712, USA}

\author[0000-0001-6004-875X]{Anna Punanova}
\affiliation{Ural Federal University, 620002 Mira st. 19, Yekaterinburg, Russia}

\author[0000-0002-0528-8125]{Elena Redaelli}
\affiliation{Max-Planck-Institut für extraterrestrische Physik, Giessenbachstrasse 1, 85748 Garching, Germany}

\author[0000-0002-9485-4394]{Samantha Scibelli}
\affiliation{Department of Astronomy, University of Arizona, Tucson, AZ 85721, USA}

\author{Yancy Shirley}
\affiliation{Steward Observatory, 933 North Cherry Avenue, Tucson, AZ 85721, USA}



\begin{abstract}

Recent observations of global velocity gradients across and along molecular filaments have been interpreted as signs of gas accreting onto and along these filaments, potentially feeding star-forming cores and proto-clusters. The behaviour of velocity gradients in filaments, however, has not been studied in detail, particularly on small scales ($< 0.1$ pc). In this paper, we present \textsc{MUFASA}, an efficient, robust, and automatic method to fit ammonia lines with multiple velocity components, generalizable to other molecular species. We also present \textsc{CRISPy}, a \textsc{Python} package to identify filament spines in 3D images (e.g., position-position-velocity cubes), along with a complementary technique to sort fitted velocity components into velocity-coherent filaments. In NGC 1333, we find a wealth of velocity gradient structures on a beam-resolved scale of $\sim 0.05$ pc. Interestingly, these local velocity gradients are not randomly oriented with respect to filament spines and their perpendicular, i.e., radial, component decreases in magnitude towards the spine for many filaments. Together with remarkably constant velocity gradients on larger scales along many filaments, these results suggest a scenario in which gas falling onto filaments is progressively damped and redirected to flow along these filaments.

\end{abstract}



\section{Introduction}\label{sect:intro}

Molecular cloud filaments appear to play a pivotal role in star formation. In addition to being featured prominently in star-forming regions (e.g., \citealt{Schneider1979}; \citealt{Bally1987}) and being ubiquitous in molecular clouds at large (e.g., \citealt{Andre2010}), filaments appear to harbour most of the observed dense cores (e.g., \citealt{Menshchikov2010}; \citealt{Konyves2015}), the smallest structure from which stellar systems emerge (see \citealt{DiFrancesco2007}). Theoretically, filaments appear to form naturally from supersonic turbulent motions of a cloud in numerical simulations, both in the absence (e.g., \citealt{Porter1994}) and in the presence (e.g., \citealt{Jappsen2005}) of self-gravity. Moreover, filaments appear to be analytically the most favoured structure to grow locally and fragment readily under weak perturbations in a finite cloud (\citealt{Pon2011}). Such properties likely make filaments highly effective at assembling dense cores from a molecular cloud prior to, or even in the absence of, an overwhelming, global cloud collapse.

How dense cores relate to their host filaments is currently not well understood. Gravitationally induced fragmentation along filament lengths, like those found in numerical models (e.g., \citealt{Bastien1991}; \citealt{Inutsuka1997}), has been suggested to be how supercritical filaments produce cores, as inferred by \textit{Herschel} observations (see \citealt{Andre2014}). While \cite{Hacar2011} found that dense structures in the L1517 filament correlate with oscillatory line-of-sight velocities, suggesting filament fragmentation, such behaviour has not been generally observed in other filaments (e.g., \citealt{Tafalla2015}). Other core formation mechanisms, such as those that form cores and filaments simultaneously in simulations (e.g., \citealt{GongOstriker2011}; \citealt{ChenOstriker2014}, \citeyear{ChenOstriker2015}; \citealt{Gomez2014}), may thus play an important role in star formation as well.

In addition to forming cores, filaments in simulations accrete material from their surroundings and transport mass along their lengths to feed dense cores and proto-clusters (e.g., \citealt{Balsara2001}; \citealt{Smith2011},  \citeyear{Smith2016}; \citealt{Gomez2014}). Indeed, velocity gradients observed across (e.g., \citealt{Palmeirim2013}; \citealt{Fernandez-Lopez2014}; \citealt{Dhabal2018}) and along (e.g., \citealt{Kirk2013}; \citealt{Friesen2013}) filaments have been interpreted as evidence for such accretion onto and along these filaments, respectively. Further kinematic studies are needed to understand how such filaments fit within the wide variety of existing models and how they assemble mass in star formation in detail.

A filament that appears to be a single, coherent (i.e., continuous) structure on the sky may not necessarily be truly coherent in three dimensional, position-position-position (ppp) space. Multiple structures that are distinct in ppp space can appear as a single structure by mere projection along lines of sight. With CO observations, \cite{Hacar2013} showed that a seemingly coherent filament on the sky can in fact contain multiple velocity-coherent `fibres' when viewed in position-position-velocity (ppv) space. 

While some ppv fibres may indeed trace physical ppp sub-filaments like those produced in simulations (e.g., \citealt{Moeckel2015}; \citealt{Smith2016}; \citealt{Clarke2017}), synthetic CO observations of a simulation showed that ppv fibres do not necessarily map well onto ppp structures, and vice versa (\citealt{Clarke2018}). Structures that are coherent in ppv space can still suffer from line-of-sight confusion when distinct ppp structures possess similar velocities (e.g., \citealt{Beaumont2013}). Such a scenario can be common, for example, when multiple ppp structures are swept up by a large-scale flow. Fortunately, denser gas tracers such as NH$_3$ and N$_2$H$^+$ are expected to be less susceptible to these problems due their lower volume-filling fraction in a cloud. This claim seems to be supported by \cite{Tafalla2015}, who observed only a single N$_2$H$^+$ ppv fibre over each line of sight where multiple CO ppv fibres had been detected earlier by \cite{Hacar2013}.

Regardless of how well ppv coherent (hereafter velocity-coherent) structures map onto ppp space, multi-component line modelling is needed to avoid deriving erroneous gas properties that are unphysical. Kinematic analyses that perform multiple-component fits to a large number of spectra, however, are uncommon. This situation is due to the typical need for human intervention in popular least-squares fitting methods, such as the Levenberg–Marquardt (LM; \citealt{Levenberg1944}; \citealt{marquardt1963}; \citealt{More1978}) method, and the inefficiencies associated with many automated approaches, such as the grid-search or Markov chain Monte Carlo (MCMC) methods.

Recent automated methods for multi-component fits, such as Behind The Spectrum (BTS; \citealt{Clarke2018}) and \textsc{GaussPy+} \citep{Riener2019}, work only with optically-thin, Gaussian lines by design. Other methods that fit hyperfine lines, such as those used by \citeauthor{Henshaw2016} (SCOUSE, \citeyear{Henshaw2016}) and \cite{Hacar2017}, are semi-automatic and hence are still subject to human biases. An efficient, automated method that fits hyperfine lines is therefore highly desirable for kinematic studies that use species like NH$_3$ and N$_2$H$^+$ to trace dense cores and filaments. 

In this paper, we describe an automated, generalizable method that fits two-component NH$_3$ (1,1) spectra efficiently using the LM method, without the need for user-provided initial guesses. The fitted models are subsequently used to identify filament spines in ppv space, which are sorted into velocity-coherent filaments accordingly. Moreover, we present a novel approach to study velocity gradients in filaments on beam-resolved scales, where velocity gradients are decomposed into components that are parallel and perpendicular with respect to local filament spines. Such a technique enables us to explore filament kinematics and accretion flow directions on the dense core ($< 0.1$ pc) scale in addition to the filament scale ($> 0.5$ pc). 

We apply our new methods to filaments seen in NH$_3$ (1,1) data of the NGC 1333 region. Located at a distance of about 295 pc away (\citealt{Ortiz-Leon2018}; \citealt{Zucker2018}), the NGC 1333 star-forming clump in the Perseus molecular cloud is one of the nearest cluster-forming regions. Its properties make NGC 1333 an ideal place to study the interplay between filaments and cores in a cluster-forming environment  in detail. NGC 1333 is also one of the most extensively studied star-forming clumps (see \citealt{Walawender2008}), providing a wealth of context within which our study can be placed. 

This paper is laid out as follows. We describe our NH$_3$ (1,1) model, synthetic data, and observed data of NGC 1333 in Section \ref{sect:data}. Methods behind our analysis, as well as test results of our line-fitting method, are presented in Section \ref{method}. The results of our analysis on the NGC 1333 observations are presented in Section \ref{results}, followed by a discussion of these results in Section \ref{discussion}. A concluding summary is in Section \ref{summary}.

\section{Models \& data}\label{sect:data}

We used two data sets for our work presented here: one synthetic and one observational. The spectral model behind our line fits is described in Section \ref{subsect:multi-slab model} while the synthetic data used to test the accuracy of our line-fitting method is described in Section \ref{subsect:synthspec}. The observations we used for this work are obtained from the Green Bank Ammonia Survey (GAS; \citealt{Friesen2017}) and are presented in Section \ref{subsect:nh3_obs}.

\subsection{NH$_3$ Line Models}\label{subsect:multi-slab model}

We modelled observations of the NH$_3$ (1,1) inversion transition along a given line of sight with up to two homogeneous bodies of beam-filling gas known as slabs. Each slab in our model corresponds to a kinematic (i.e., velocity) component observed in a spectrum and is assumed to be at a local thermal equilibrium with itself. Furthermore, we assume the emission can be parameterized by excitation temperature ($T_{ex}$), optical depth ($\tau_{0}$), velocity dispersion ($\sigma_{v}$), and velocity centroid in the local standard of rest frame ($v_{\textup{LSR}}$). The optical depth profile of each slab is described by
\begin{equation} \label{eq:tau_hyperfines}
\tau(v, \sigma_{v}) = \sum_{i=1}^{n} W_i\tau_0 \exp{\left[ \frac{-(v-v_{\mathrm{LSR}} -\delta v_i)^2}{2\sigma_v^2}\right]},
\end{equation}
where $W_i$ and $\delta v_i$ are the relative weight and velocity offset of each of the eighteen NH$_3$ (1,1) hyperfine components, respectively. These weights and velocity offsets are tabulated by \cite{Mangum2015}. We further note that the $\tau_{0}$ here corresponds to the combined optical depth of all the hyperfine components. 

The radiative transfer of our model emission through each slab is governed by:
\begin{equation} \label{eq:rad_trans_sol}
I_\nu = B_\nu(T_{ex})(1-e^{-\tau_\nu})+I^{bg}_\nu e^{-\tau_\nu},
\end{equation}
where $I^{bg}_\nu$ is the specific intensity of the background radiation and $B_{\nu}(T)$ is the Planck function at a temperature $T$. Each slab is assumed to have a constant $T_{ex}$ and we adopt the cosmic microwave background (CMB), as the $I^{bg}_\nu$ for our first slab, i.e., the slab furthest from the observer. We then subsequently use the emergent $I_\nu$ of the first slab as the $I^{bg}_\nu$ of our second slab to complete the calculation. To mimic baseline-removal used in our data reduction, a constant value of $B_{\nu}(T_{\mathrm{CMB}}= 2.73 \mathrm{K})$ is subtracted from Equation \ref{eq:rad_trans_sol} in our final $I_{\nu}$ model.

\begin{figure*}[t!]
\centering
\includegraphics[width=0.46\textwidth]{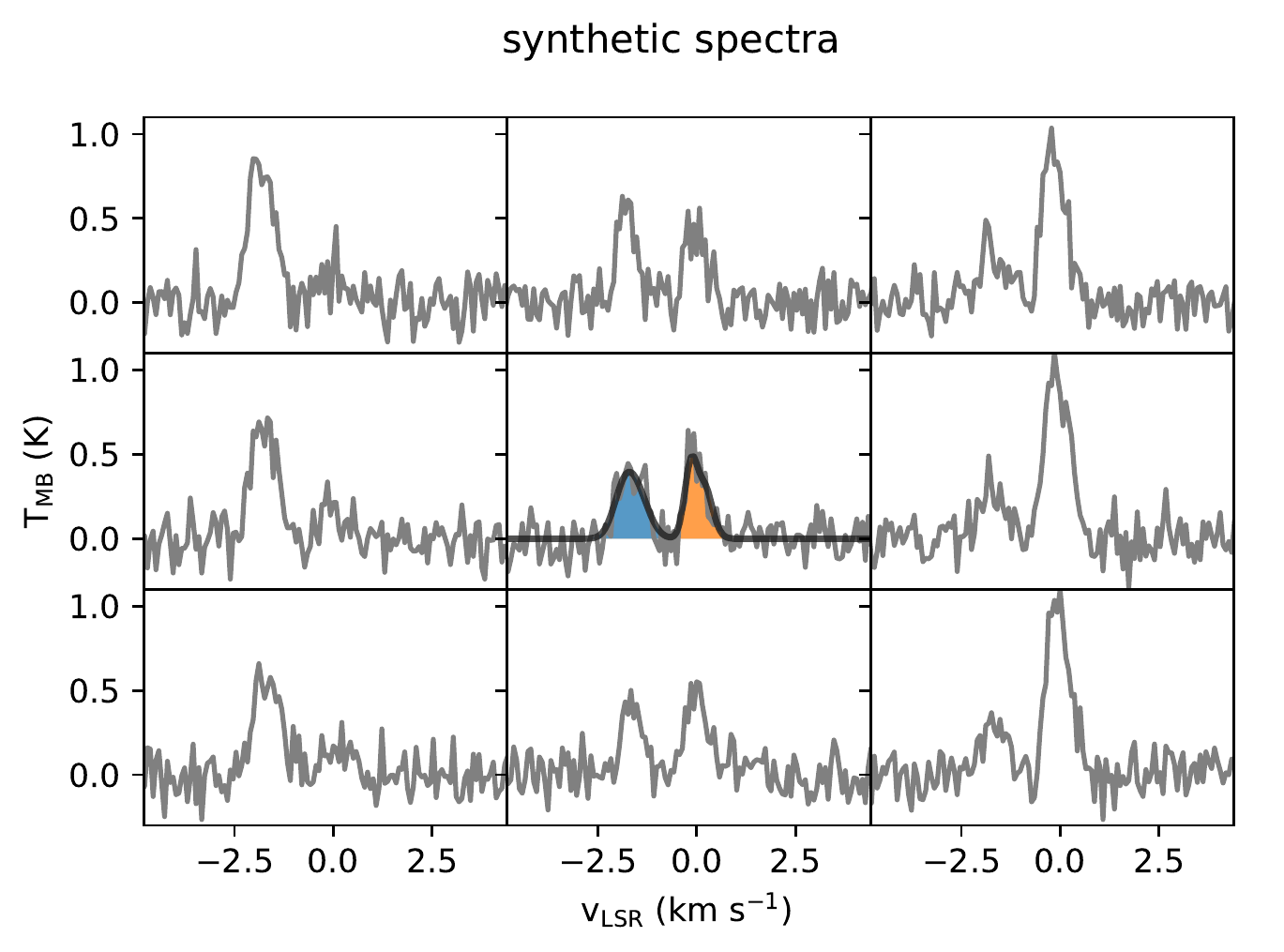}
\hspace{7mm}
\includegraphics[width=0.46\textwidth]{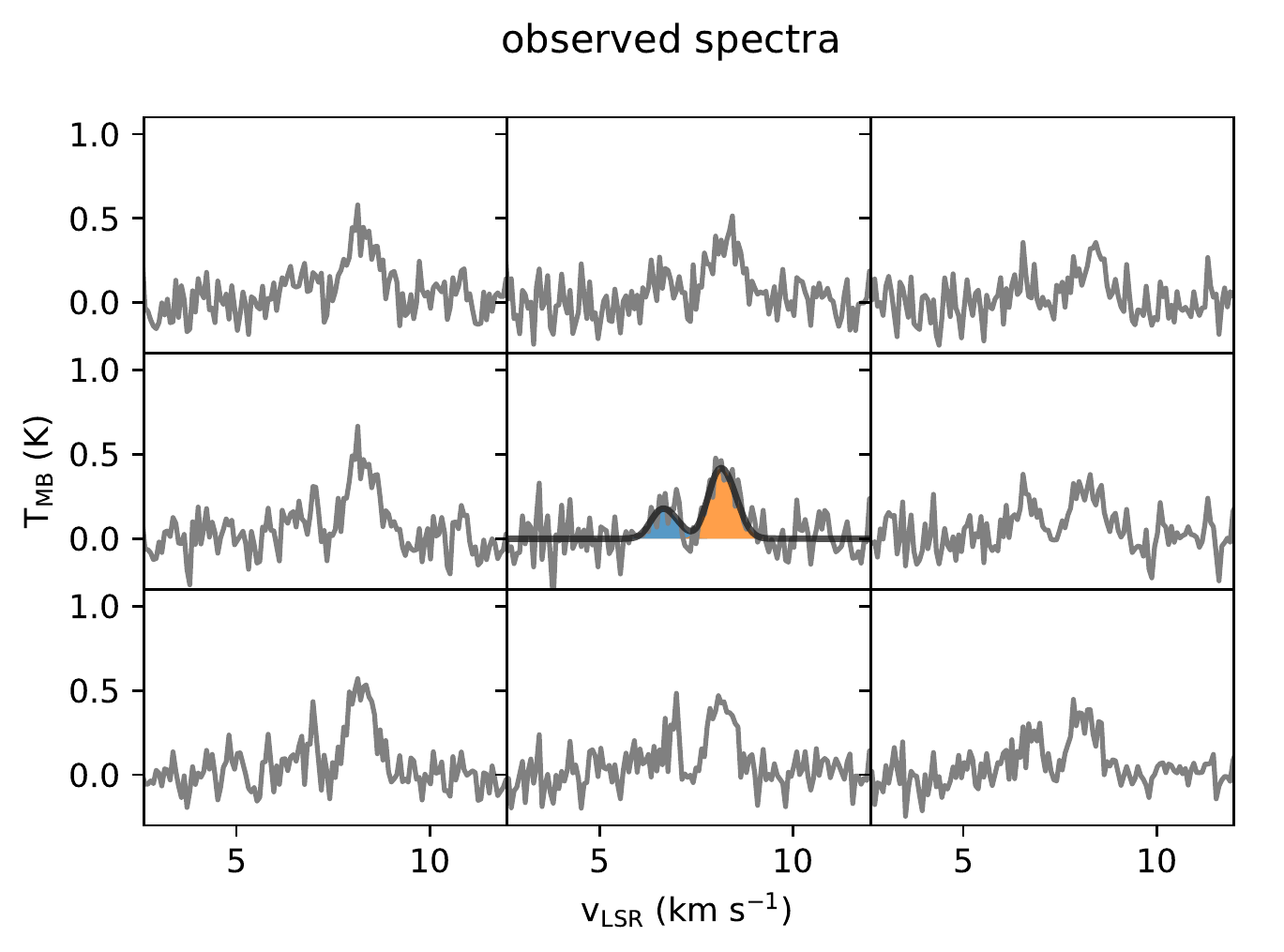}
\caption{NH$_3$ (1,1) spectra (grey) taken from a synthetic spectral cube (left) and the NGC 1333 observation (right) over a grid of $3\times3$  footprint, zoomed in to focus on the main hyperfine structures. Models fitted to all 18 hyperfine component of the spectra in the center pixel are shown in black, and the individual components of the model are shown in blue and orange.}
  \label{fig:spec_eg}
\end{figure*}

While we assume the slab furthest from an observer to be the optically thicker slab in our initial guesses, our least-squares fitting routine ultimately decides the order of the modelled slabs along the line of sight. We note that such an ordering of the slabs is unimportant when the two slabs are optically thin with respect to each other, or when two optically thick slabs are not spectrally overlapped due to large $v_{\mathrm{LSR}}$ separations. Since the satellite hyperfine lines of NH$_3$ (1,1) emissions are optically thin in most cases, they help to constrain our kinematic fits well even when the main hyperfine lines are optically thick. Such a constraint allows our fitting method to distinguish double spectral peaks resulting from a line absorption profile from those resulting from superpositions. We implicitly explore how the ordering of the slabs affects our fitting accuracy with the performance test described below in Section \ref{subsect:synthspec}. 

\edit3{Since this paper focuses exclusively on understanding the gas kinematics, we analyzed only the NH$_3$ (1,1) lines to maximize our spatial coverage for the study. As the (2,2) line is expected to be weaker than the (1,1) line, this strategy enabled us to extend our analysis over wider regions where the NH$_3$ (2,2) line is not detected. While such an approach does not allow us to derive $T_{ex}$ and $\tau_0$ accurately for purely optically thin slabs, where the two parameters are spectrally degenerate, it does remove the potential bias that comes with assuming a single $T_{ex}$ for both the (1,1) and (2,2) lines.}

\subsection{Synthetic Spectra}\label{subsect:synthspec}

\edit3{To test the accuracy of our line-fitting method, we generated a set of 25,000 synthetic NH$_3$ (1,1) spectra, each from either a one- or two-slab model that fills the beam. To motivate our test set physically, we constructed each of our synthetic slabs with the NH$_3$ model adopted by the GAS DR1 first results paper (\citealt{Friesen2017}) instead of using our fitting model, described earlier in Section \ref{subsect:multi-slab model}. The model adopted by the DR1 paper is based on the work by \cite{Rosolowsky2008} and \cite{Friesen2009}, and developed from the framework described in \cite{Mangum2015}.}

\edit3{Each gas slab in our testing model is physically parametrized by the para-ammonia column density, $N$, and the kinetic temperature, $T_{\mathrm{k}}$, in addition to $\sigma_{v}$, $v_{\textup{LSR}}$, and $T_{ex}$. We further assumed these slabs are in local thermal equilibrium (LTE), i.e., $T_{ex} = T_{\mathrm{k}}$, and draw parameters behind each instance of synthetic slab randomly from the pre-defined distributions specified below:}

\begin{itemize}  
\item $N$ is drawn from a log-uniform distribution in the range of $13 \leq \log{(N/\mathrm{cm}^{-2})} < 14.5$;
\item $T_{\mathrm{k}}$ is drawn from a uniform distribution in the range of 8 K $ \leq T_{\mathrm{k}} < $ 25 K;
\item $\sigma_{v}$ is the quadrature sum of the thermal line width, $\sigma_{v, \mathrm{T}} = 0.08$ km s$^{-1}$, and the non-thermal line width, $\sigma_{v, \mathrm{NT}}$, where $\ln{ [ \sigma_{v, \mathrm{NT}}/(\mathrm{km \ s}^{-1}) ]}$ is drawn from a normal distribution with a mean and a standard deviation of -2.3 and 1.5, respectively. This particular line width distribution is chosen to resemble those found in the GAS DR1 first results, and;
\item $v_{\textup{LSR}}$ of the first gas slab is drawn from a uniform distribution in the range of -2.5 km s$^{-1} \leq v_{\textup{LSR}} <$ 2.5 km s$^{-1}$ while the $v_{\textup{LSR}}$ offset of the second slab from the first is drawn independently from the same distribution.
\end{itemize}
\edit3{We chose these distributions to represent broadly the typical physical conditions seen towards nearby molecular clouds with NH$_3$.}

We constructed a two-slab spectrum using the same radiative transfer formalism described by Equation \ref{eq:rad_trans_sol}, with the CMB subtracted as a constant. The final synthetic spectrum is produced by adding random Gaussian noise with an rms value of 0.1 K to the model spectrum. The value of 0.1 K is chosen to mimic the typical noise level found in the GAS DR1 observations \citep{Friesen2017}.

For each instance of randomly generated spectrum, eight additional, spatially-correlated spectra are produced. Collectively, these spectra are placed in a 3$\times$3 spatial grid with the original spectrum positioned in the centre. The purpose behind creating such a cube is to provide spatial information for fitting methods that utilize them, including the method presented in this paper. 

The spatial correlation between pixels of a synthetic cube is achieved by applying spatial gradients to the model parameters, referenced at the central pixel. The gradients used for each parameter are randomly drawn from a Gaussian distribution with a 1-$\sigma$ values of 0.2 K, 0.1 km s$^{-1}$, 0.1 km s$^{-1}$, and 0.01 for $T_{\mathrm{k}}$, $\sigma_{v}$,  $v_{\textup{LSR}}$, and $\log{(N)}$, respectively, per pixel. Figure \ref{fig:spec_eg} (left) shows spectra extracted from such a synthetic cube, displayed spatially on a $3\times3$ grid.


\subsection{NH$_3$ Observations, Reduction, and Imaging}\label{subsect:nh3_obs}

We observed the Gould Belt molecular clouds with NH$_3$ (1,1) and (2,2) inversion lines as a part of the GAS survey (\citealt{Friesen2017}). The GAS observations were made with the Robert C. Byrd Green Bank Telescope (GBT) using its 7-beam K-Band Focal Plane Array (KFPA) and its VErsatile GBT Astronomical Spectrometer (VEGAS) backend. The angular and spectral resolutions (FWHMs) of our NH$_3$ data are 32$''$ and $\sim 0.07$ km s$^{-1}$ (i.e., 5.7 kHz at 23.7 GHz), respectively.

Our targets are observed using the On-The-Fly (OTF) technique, where a $10' \times 10'$ on-sky tile was scanned with a Nyquist-sampled spacing between each row. We reduced these observations with the GBT KFPA data reduction pipeline (\citealt{Masters2011}) and imaged them with the recipe described by \cite{Mangum2007}. Further details of the observations, data reduction, and imaging are available in the GAS first results paper (\citealt{Friesen2017}). For this work, we will focus on the observations of NGC 1333 star-forming clump, available to the public via Data Release 1 (DR1) of the first results paper. Figure \ref{fig:spec_eg} shows example spectra on its right panel extracted from the DR1 data over a $3 \times 3$ pixel region in NGC 1333.

\section{Analysis Methods}\label{method}

Here, we present our analysis methods in this section below. Details of our spectral fitting method are provided in Section \ref{subsect:multi-slab fitting}. We conducted a performance test on our fitting method to quantify its accuracy and completeness, and we present the details and the results of this test in Section \ref{sub:linefit_tests}. Methods for identifying velocity-coherent filaments and assigning component memberships to these filaments based on the fits are presented in Section \ref{subsect:vcFilID}. Methods behind velocity gradient analysis are presented in Section \ref{subsect:VGradAnalysis}.

\subsection{Spectral Fitting}\label{subsect:multi-slab fitting}

We fitted our synthetic and real data automatically using the Levenberg–Marquardt (LM; \citealt{Levenberg1944}; \citealt{marquardt1963}; \citealt{More1978}) least-squares minimization method. We bypassed the need for user-provided initial guesses using an automated approach described in Section \ref{subsub:multi-slab fitting} and performed least-squares fits using the \textsc{PySpecKit} package \citep{Ginsburg2011}. The \texttt{Python} implementation of the LM method used by the \textsc{PySpecKit} is based on the \texttt{FORTRAN} version found in the \textsc{MINPACK-1} package \citep{More1980}, made available via a series of translations (\citealt{Rivers2002}; \citealt{Markwardt2009}). Our fits are performed on a pixel-by-pixel basis for all the pixels in our data, excluding noisy regions near the map edges. We use a statistical method described in \ref{subsub:mod_selection} to discern whether a pixel is better modelled by noise, a one-component model, or a two-component model. \edit1{Our fitting package is publicly available via \textit{GitHub} under a GNU General Public License as the \textsc{MUFASA}\footnote{\texttt{MUFASA} codebase: \url{https://github.com/mcyc/mufasa}.} (i.e., MUlti-component Fitter for Astrophysical Spectral Applications). The version we used for this work is archived in Zenodo \citep{MUFASADOI}.}

\subsubsection{Making Initial Guesses}\label{subsub:multi-slab fitting}

The LM method is an iterative approach to find local minima using a hybrid algorithm of the gradient-ascent and the Gauss–Newton methods (see \citealt{Lourakis2005}). Due to the nature of these methods, good initial guesses are typically needed with the LM method to find the global minima. Such guesses are particularly required for complex spectral models with many local chi-squared minima and are the reasons why many earlier efforts to fit multiple slab models (e.g., \citealt{Hacar2013}; \citealt{Henshaw2016}) require human intervention and are not fully automated. For our method, i.e. \textsc{MUFASA}, we fit spectra automatically using initial guesses generated from a recipe described below. A statistical technique (see Section \ref{subsub:mod_selection}) is used thereafter to decide which of the one- or two-slab models fitted for each pixel is more appropriate, without human intervention.

\begin{figure*}
\centering
\includegraphics[width=0.9\textwidth]{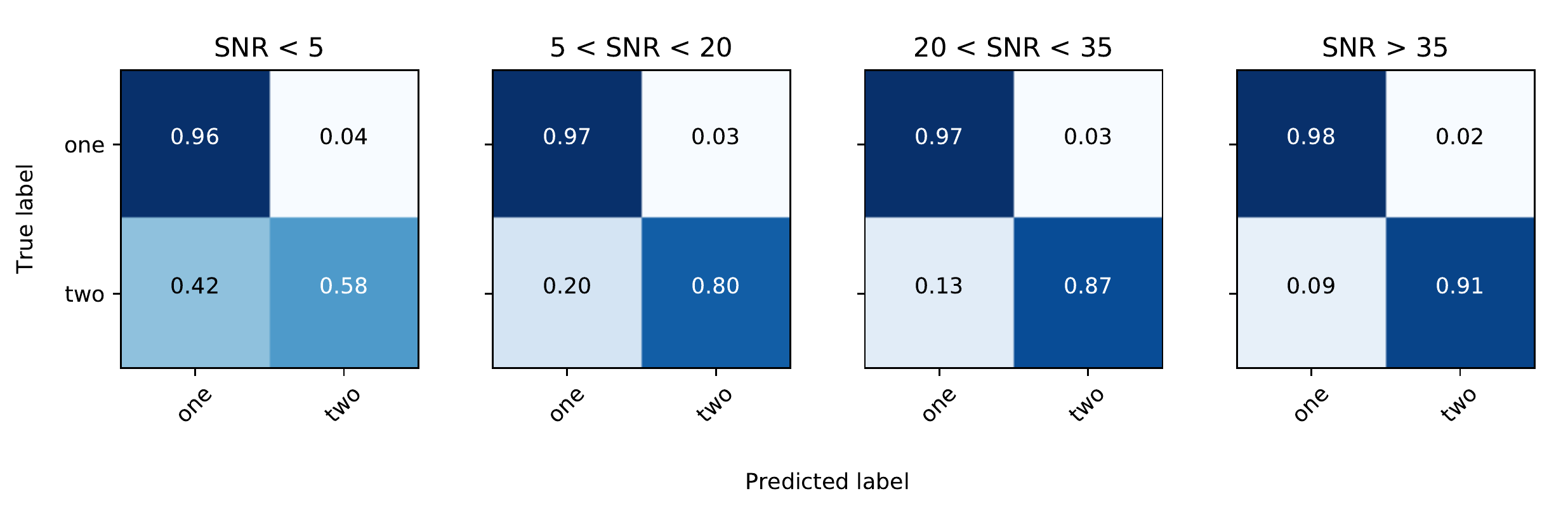}
\caption{Confusion matrices quantifying \textsc{MUFASA}'s ability to classify the number of velocity slabs behind a synthetic spectrum. Each matrix contains a subset of our samples at a given range of SNR values.}
  \label{fig:cmatrix}
\end{figure*}

The GAS DR1's one-slab fitting method used an effective and automated recipe to make initial guesses. The recipe used the first and second moments of the central, i.e., main, NH$_3$ (1,1) hyperfine lines as its initial $v_{\mathrm{LSR}}$ and $\sigma_{v}$ guesses. Such a calculation excludes the satellite hyperfine lines because their large velocity offsets are not a kinematic feature. The main hyperfine lines were isolated for such a calculation in DR1 via a user-defined spectral window.

For our one-slab model fits, we adopt the DR1 recipe for our initial guesses of $v_{\mathrm{LSR}}$ and $\sigma_{v}$, \edit3{but define our spectral windows automatically instead. We automate such a process by centering a 6 km s$^{-1}$ full-width window on the emission peak of the spatially integrated spectrum of our data. Since the NH$_3$ (1,1) emission should be optically thin throughout the majority of the data, such an emission peak should locate the whereabouts of the main hyperfine lines. Once the window is defined, we follow the DR1 recipe and calculate the zeroth, first, and second moments ($\mu_0$, $\mu_1$, $\mu_2$) of the main hyperfine lines over the window.}

To obtain initial guesses for $T_{ex}$ and $\tau_0$ better than assuming fixed values, we use the zeroth moment ($\mu_0$) map as a proxy instead. Specifically, we calculate our guesses by first normalizing the 99.7 percentile value of the $\mu_0$ distribution across the map to one. The initial guesses for $T_{ex}$ and $\tau_0$ are then obtained from the normalized $\mu_0$, i.e., $\widetilde{\mu_{0}}$, as $\widetilde{\mu_{0}} \cdot T_{gmx}$ and $\widetilde{\mu_{0}} \cdot \tau_{gmx}$, respectively, where $T_{gmx} = 8$ K and $\tau_{gmx}= 1$. The resulting fits from adopting such guesses do not depend sensitively on $T_{gmx}$ and $\tau_{gmx}$ around these chosen values. 

We expand the DR1 fitting recipe further for our two-slab fits via the following steps:

\begin{enumerate}  
\item Adopt the first and second $v_{\mathrm{LSR}}$ guesses as $\mu_1 \pm 0.4 \ \mu_2$, respectively.
\item Adopt the first and second $\sigma_{v}$ guesses both as $0.5 \ \mu_2$, respectively.
\item Adopt the first and second $T_{ex}$ guesses as $\widetilde{\mu_{0}} \ T_{gmx}$  and $0.8 \ \widetilde{\mu_{0}} \ T_{gmx}$, respectively.
\item Adopt the first and second $\tau_0$ guesses as $0.75 \ \widetilde{\mu_{0}} \ \tau_{gmx}$ and $0.25 \ \widetilde{\mu_{0}} \ \tau_{gmx}$, respectively.
\end{enumerate}

As with the guesses used for one-slab fits, our choices for $T_{gmx}$ and $ \tau_{gmx}$, and their respective scaling factors, do not affect the fitting outcome sensitively. Our choices for these values are motivated by a hypothetical scenario where the two gas slabs emitting a spectrum have comparable velocity dispersions, densities, and kinetic temperatures. We note that our initial guesses assume the slab further from the observer has a higher optical depth. As we show below in Section \ref{sub:linefit_tests}, our recipe for making guesses for the two-slab fit is robust even when the gas slabs emitting the spectrum have dissimilar velocity dispersions, i.e., contrary to this assumption. 

To take advantage of spatial information present in our observations, we first fit data that are spatially convolved to an angular resolution twice the size of the original resolution. The parameter maps derived from this initial fit are then median-smoothed and interpolated. For $T_{ex}$ and $\tau_{0}$ guesses, values outside of ranges 3 - 8 K and 0.2 - 8, respectively, are removed prior to median-smoothing and interpolation. These post-processed parameter maps are then adopted as the initial guesses of our fits to the full resolution cubes. 

By using parameters fitted to the spatially convolved cube as initial guesses for the full-resolution fit, we are able to take advantage of the enhanced signal-to-noise ratio (SNR) in addition to the spatial information present in the convolved cube to improve our fits. Figure \ref{fig:spec_eg} shows spectra extracted over a $3 \times 3$ pixel region from a synthetic data cube (left) and the DR1 data of NGC 1333 (right), demonstrating the spatial correlation of these spectra between pixels. The two-slab model fitted to the central pixel using this method is overlaid over the central spectrum.

Since moment estimates for making initial guesses may overlook a faint spectral component in the presence of a much brighter one, we further fit one-slab models to our fit residuals in an attempt to recover missing components. The spectral window used to estimate initial guesses for this fit is centered on the $v_{\mathrm{LSR}}$ derived from the initial one-slab fit, with a full window width of 7 km s$^{-1}$. 

We use the results of the one-slab residual fit subsequently to assist with the recovery of a missing component. These results are taken in tandem with those of the original one-slab fit as initial guesses for the re-attempt at fitting a two-slab model. We perform these re-attempts over pixels where one-slab models initially fit the full spectrum better than the two-slab models, as determined by the selection criterion described in the next section (i.e., Section \ref{subsub:mod_selection}). The same criteria is used further to determine whether or not this new two-slab fit is justified over the one-slab fit.

\subsubsection{Model selection}\label{subsub:mod_selection}

\begin{figure*}[t!]
\centering
\includegraphics[width=0.46\textwidth]{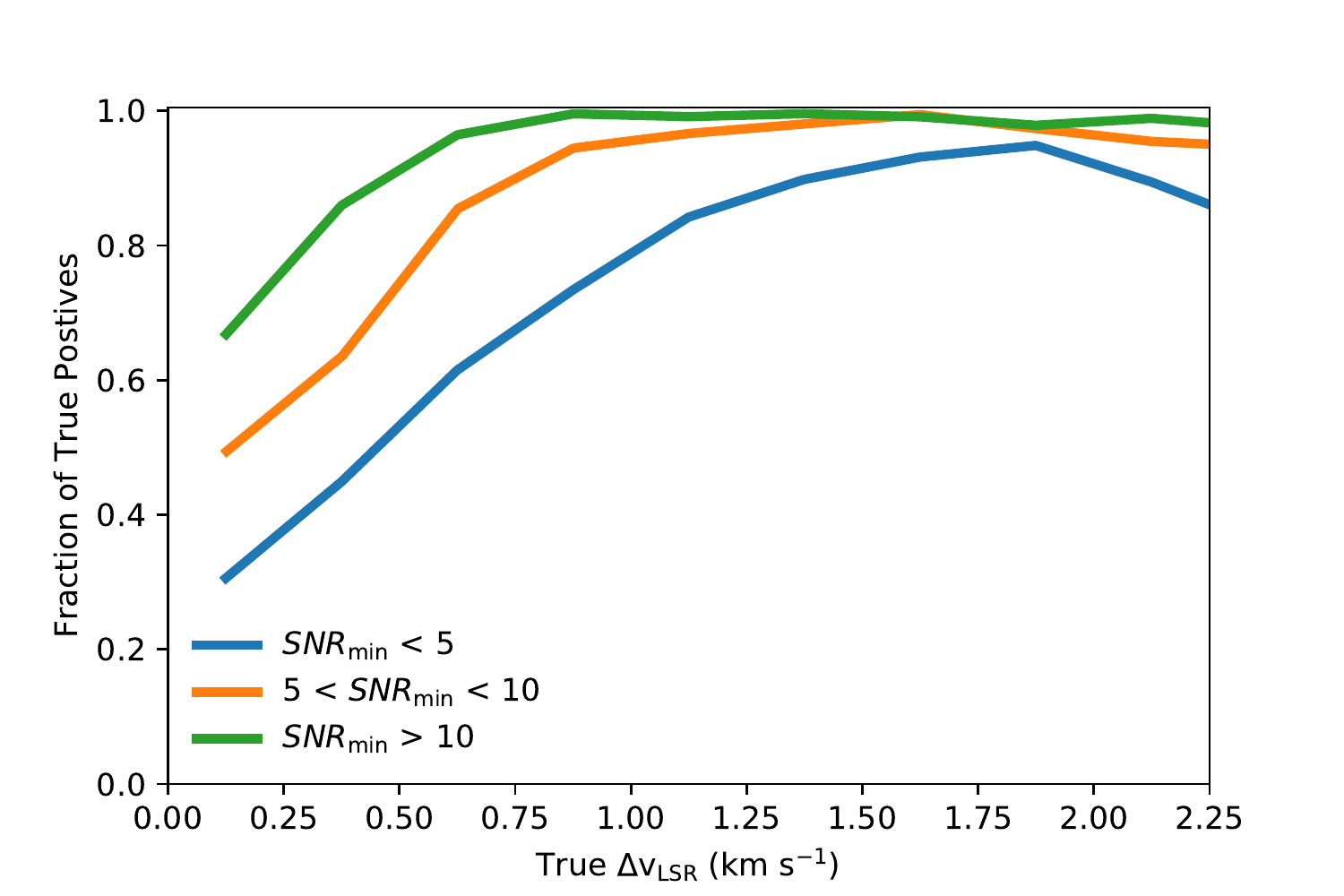}
\hspace{6mm}
\includegraphics[width=0.46\textwidth]{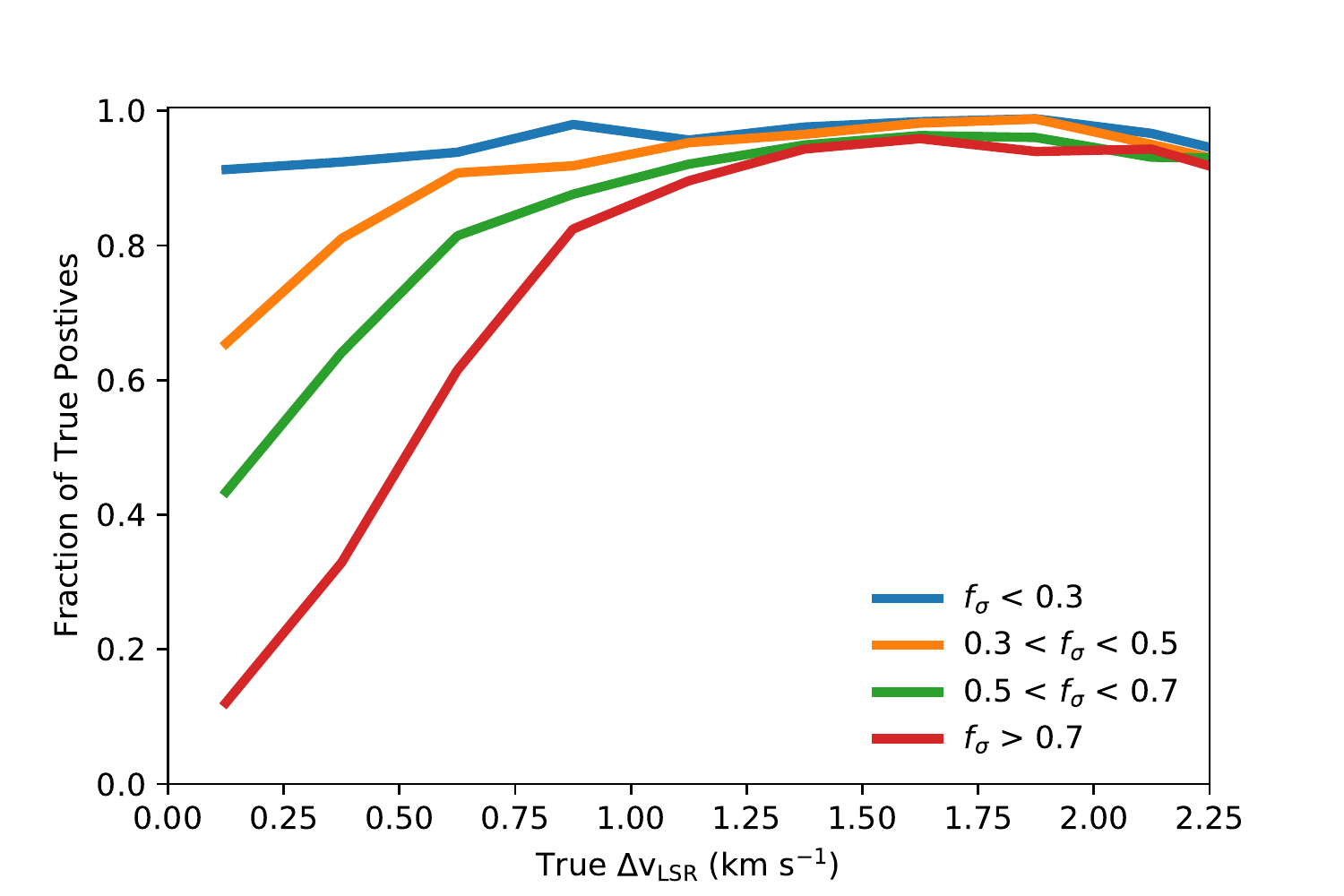}
\caption{The rate of true-positive identification of a two-slab spectrum plotted as a function of the $v_{\mathrm{LSR}}$ separation between the slabs. The left panel shows these test samples binned by the SNR values of the fainter slab while the right panel shows these values binned by velocity dispersion ratio $f_{\sigma}$ of the narrow slab with respect to the wide slab. The median $\sigma_v$ values for the narrow and wide components in these samples are 0.23 km s$^{-1}$ and 0.48 km s$^{-1}$, respectively.}
  \label{fig:2compID_vSep}
\end{figure*}

Many previous multi-component fitting methods selected their best fit models for each component via an SNR threshold (e.g., \citealt{Sokolov2017}), a velocity separation threshold, or both (e.g., \citealt{Hacar2013}, \citealt{Hacar2017}, \citealt{ChenV2019}). While these criteria are effective at avoiding overfitting, they are not necessarily good at picking up all the spectral components present along a given line of sight. We addressed this issue by using the corrected Akaike information criterion (AICc; \citealt{Akaike1974}; \citealt{Sugiura1978}) to select the best model between the one- and two-slab fits, on a pixel-by-pixel basis. 

The AICc is a second-order corrected estimator, based on the K-L information loss (\citealt{Kullback1951}), for the relative likelihood of one model with respect to another at representing a dataset with $N$ samples. Assuming the errors in the data are normally distributed, the AICc can be written in terms of the $\chi^2$ of the fit as:

\begin{equation} \label{eq:AICc}
\mathrm{AICc} = \chi^2 + 2p + \frac{2p(p+1)}{N-p-1},
\end{equation}
where $p$ is the number of parameters used in the model. At each pixel, we accept a two-slab model as the better fit over its one-slab counterpart when their relative likelihood, $K^2_1$, given by 
\begin{equation} \label{eq:lnk21}
\ln{K^2_1} = \exp{(\mathrm{AICc}_{2} - \mathrm{AICc_{1})}}
\end{equation}
is greater than 5 \citep{Burnham2004}. Similarly, we accept a one-slab model as the better fit over a $I_{\nu} =0$ model of a noise with no free parameter when their relative likelihood, $\ln{K^1_0}$, is greater than 5.

Pixels with reduced $\chi^2$ values of $\chi_{\nu}^2 > 1.5$ are further masked from our analysis to ensure spectra which are inadequately modelled by our fits, e.g., those with possibly three or more velocity slabs, are not included in the analyses. In NGC 1333, no pixel is masked out as all pixels best fitted by 2-slab models have $\chi_{\nu}^2 < 1.5$.

\subsection{Performance Tests on Line Fitting}\label{sub:linefit_tests}

Figure \ref{fig:cmatrix} summarizes the performance of our line-fitting method, \textsc{MUFASA}, at identifying the correct number of velocity slabs behind a synthetic spectrum, using confusion matrices. These results are obtained from fits to all 25,000 synthetic test spectra described in Section \ref{subsect:synthspec} and are binned into separate matrices according to the SNR of their true underlying spectra. Unless stated otherwise, we refer to SNR as the modelled peak brightness to noise ratio of a final, radiatively transferred spectrum rather than that of its individual components. As illustrated in Figure \ref{fig:cmatrix}, our fitting method identifies true one-slab spectra robustly. The true-positive rate of this identification is greater than 96\% even for our lowest SNR ($<5$) bin. For two-slab identification, the true-positive rate correlates with the SNR value of the spectra, reaching values upwards of about 90\% at higher SNR. Even at moderate SNRs between 5 and 20, the true-positive rate is roughly 80\% for two-slab spectra. 

We next explore the impact of velocity separation between the slabs and intrinsic velocity dispersion (i.e., $\sigma_{v}$) on the success of our fitting method. Figure \ref{fig:2compID_vSep} shows the rate of true-positive identification of a two-slab spectrum as a function of the $v_{\mathrm{LSR}}$ separation between the slabs, i.e., $\Delta v_{\mathrm{LSR}}$. The left panel shows these rates binned according to the SNR value of the fainter slab, i.e., SNR$_{\mathrm{min}}$. 

The response curves in Figure \ref{fig:2compID_vSep} show the same shape across various SNR$_{\mathrm{min}}$ regimes. Namely, they increase monotonically with $\Delta v_{\mathrm{LSR}}$ until the fraction of true-positive identification plateaus at 100\%. Prior to plateauing, these curves shift vertically upwards as the SNR$_{\mathrm{min}}$ increases and behave qualitatively the same even when they are binned instead according to the SNR of the brighter slab, i.e., SNR$_{\mathrm{max}}$, or the SNR taken from the peak of the combined spectrum (i.e., the SNR).

\begin{figure*}
\centering
\includegraphics[width=0.46\textwidth]{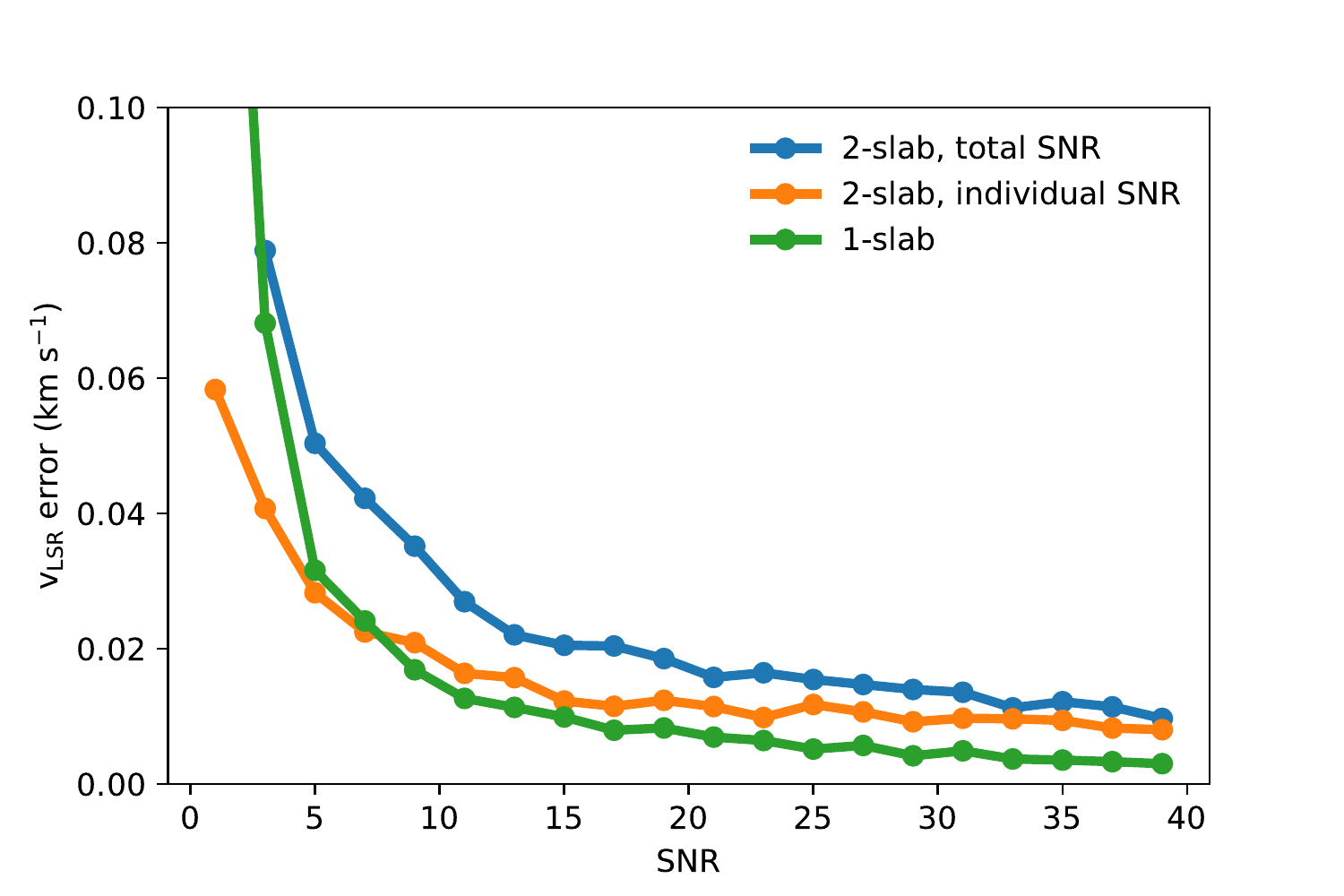} 
\hspace{6mm}
\includegraphics[width=0.46\textwidth]{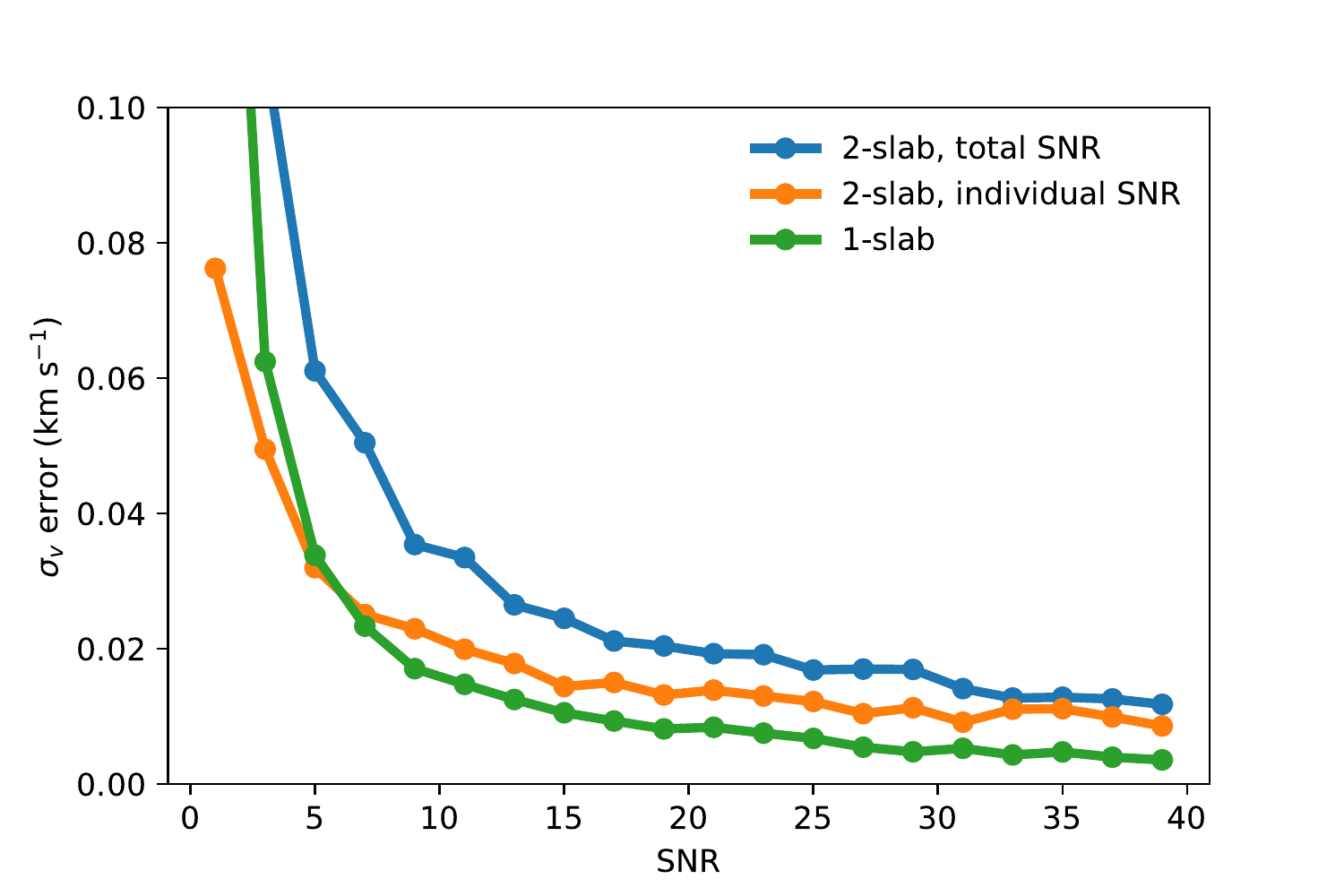}
\caption{The 1-$\sigma$ error of the fitted $v_{\mathrm{LSR}}$ (left) and $\sigma_v$ (right) as a function of (blue) the total SNR of a 2-slab spectrum, (orange) the SNR of a given slab in a spectrum, and (green) the SNR of a one-slab spectrum, for fits with correct number of identified components.}
  \label{fig:eVlsr}
\end{figure*}

The only exception to this plateauing trend is when SNR$_{\mathrm{min}}$ is low, i.e., at a value less than 5. At this low SNR regime, the true-positive rate reaches a maximum of 95\% at a value of about 1.9 km s$^{-1}$ before it turns around instead of plateauing. This turnaround in the true-positive rate likely reflects the limit of our ability to make initial guesses, suggesting that moment maps may have trouble picking up fainter components when these components reside near the edges of the spectral window from which moment maps are calculated from.

The performance of \textsc{MUFASA} decreases with decreasing $\Delta v_{\mathrm{LSR}}$, likely resulting from a lack of velocity acuity between two slabs when they have similar velocities. As $\Delta v_{\mathrm{LSR}}$ decreases, the spectral profiles of the two slabs start to blend together, making them more difficult to distinguish from a one-slab profile. This lack of acuity is what prompted many studies to adopt a $\Delta v_{\mathrm{LSR}}$ threshold for their model selection to guard against overfitting (e.g., \citealt{Hacar2013}), and remains a challenge even for advanced machine learning techniques (e.g., \citealt{Keown2019}). 

The right panel of Figure \ref{fig:2compID_vSep} shows the true-positive identification rates divided into various regimes of velocity dispersion ratio, i.e., the line width of the narrower slab relative to its wider counterpart along a line of sight ($f_{\sigma} = \sigma_{\mathrm{v, narrow}}/\sigma_{\mathrm{v, wide}}$). Here, the true-positive rate anti-correlates strongly with $f_{\sigma}$. This true-positive rate is likely enhanced when the spectral profiles from the slabs are less similar, i.e., when $f_{\sigma}$ is low, which makes them easier to discern from one another when their amplitudes are different. Indeed, this rate can be as high as 90\% for spectra with $f_{\sigma} < 0.3$ and correlates weakly with $\Delta v_{\mathrm{LSR}}$ in this regime. 

Such enhanced identification rates in the low $f_{\sigma}$ regime make \textsc{MUFASA} particularly useful for disentangling subsonic gas from supersonic gas along lines of sight. For reference, the median $\sigma_v$ values for the narrow and wide components in our test samples are 0.23 km s$^{-1}$ and 0.48 km s$^{-1}$, respectively, whereas the isothermal sound speed at 10 K is $\sim 0.2$ km s$^{-1}$. Like the trend seen in the left panel, the true-positive rate correlates with $\Delta v_{\mathrm{LSR}}$ for most $f_{\sigma}$ values prior to plateauing. The slope of these correlations, however, becomes shallower as $f_{\sigma}$ decreases.

Figure \ref{fig:2compID_vSep} reveals that \textsc{MUFASA} is able to recover a large fraction of two-slab spectra that would otherwise be missed by using a $\Delta v_{\textrm{LSR}}$ threshold for model selection. For a spectral population described by our synthetic spectra, at least 20\% and 30\% of the second slab missed by a threshold of 0.25 $\mathrm{km \ s}^{-1}$ (e.g., \citealt{Hacar2017}) and a 0.4 $\mathrm{km \ s}^{-1}$ (e.g., \citealt{ChenV2019}), respectively, are recovered with \textsc{MUFASA}. These recovery rates can be significantly higher, however, depending on the SNR and $f_{\sigma}$.

Figure \ref{fig:cmatrix} reveals that \textsc{MUFASA} rarely overfits one-slab spectra, i.e., misidentifying a one-slab spectrum as having two slabs. Our test shows \textsc{MUFASA} only misidentifies one-slab spectra $< 3$\% of the time. Even in the lowest SNR regime, such misidentification only occurs 4\% of the time. Performing model selection via AICc alone is thus sufficient to guard against overfitting without needing an additional threshold criterion.

To quantify how well \textsc{MUFASA} captures the true $v_{\mathrm{LSR}}$ and $\sigma_v$, Figure \ref{fig:eVlsr} shows the true 1-$\sigma$ errors of these two parameter fits as a function of the following values: the SNR of a two-slab spectrum, the SNR of an individual slab in a two-slab spectrum, and the SNR of a one-slab spectrum. These SNR-error relations are plotted with errors calculated from the median absolute deviation (MAD) of the fitted parameters with respect to the true value behind a synthetic spectrum. As expected, the parameter errors are anti-correlated with the SNR. We note that these errors also account for any potential cases where the spatial order of velocity slabs may not be ordered correctly along a line of sight.

\subsection{Identifying Velocity Coherent Filaments}\label{subsect:vcFilID}

\begin{figure*}
\centering
\includegraphics[width=0.75\textwidth]{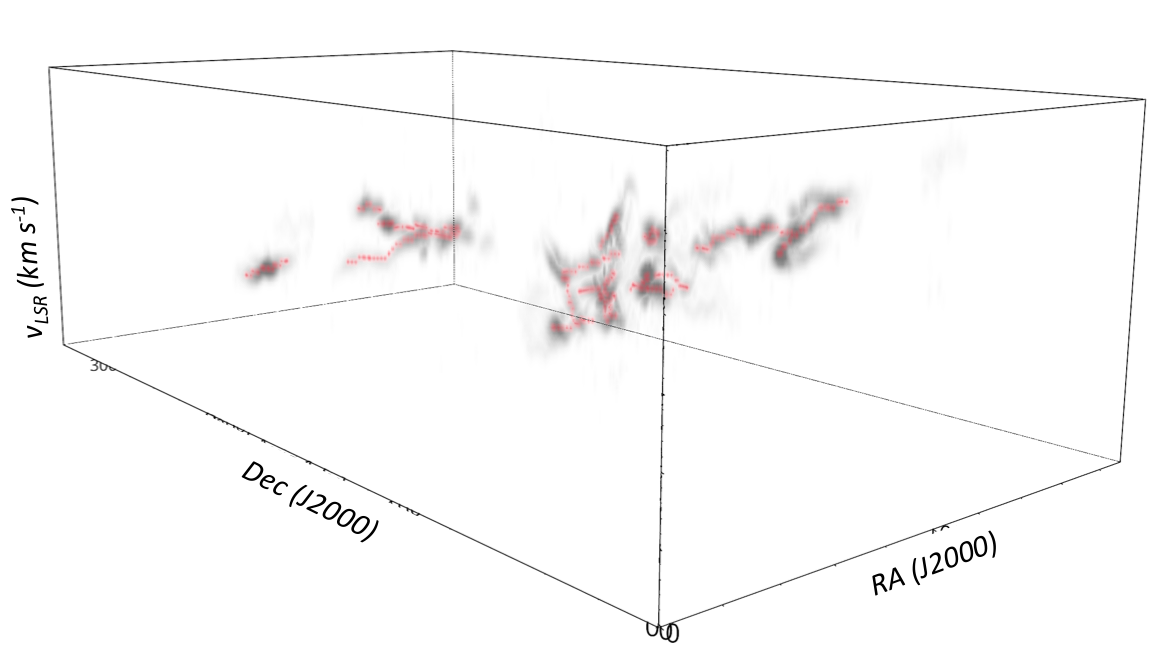}
\caption{The debelended ppv cube of NGC 1333 reconstructed from the best fit models (grey), and the filament spines identified from the cube using \textsc{CRISPy} (red).} 
  \label{fig:ppvCubeAndSpine}
\end{figure*}

Velocity gradient analyses are only useful for understanding gas flows if the structures being analyzed are velocity-coherent, i.e., continuous in velocity. Velocity slabs derived from our fits must therefore be sorted into velocity-coherent structures prior to such analyses. In this sub-section, we present methods to reconstruct fits to our data as simple emission line models in position-position-velocity (i.e., ppv) space without hyperfine structures. We further present methods to identify filament spines from these models in ppv space and sort the fitted slabs into velocity-coherent filaments based on these identified spines.

\subsubsection{Reconstructing velocity structures}\label{subsub:deblending}

To help identify velocity coherent structures in ppv space, we first reconstruct our best fit models with the hyperfine structures removed to avoid confusion from these non-kinematic features.  Such a reconstruction, known as ``deblending,'' is accomplished by computing the spectral profile of each gas slab using a single Gaussian $\tau_{\nu}$ profile based on our best fit model, which accounts for all 18 hyperfine structures (see Section \ref{subsect:multi-slab model}). In other words, we reconstruct the spectral profile along a line of sight with either one or two Gaussian $\tau_{\nu}$ components using our best fit parameters and number of components as determined by the AICc criterion.

Since our $\tau_0$ value derived from a fit (see Eq. \ref{eq:tau_hyperfines}) represents the peak optical depth of all the NH$_3$ (1,1) hyperfine lines combined, we scale down our fitted $\tau_0$ by a factor of 10 to represent better the \edit3{actual optical depths} of individual hyperfine groups in our reconstruction. For reference, the main and satellite hyperfine groups each contain about 50\% and 10\% of the optical depth represented by $\tau_0$, respectively. The observed satellite lines are thus typically optically thin even when the main hyperfine lines are not, \edit3{which enables these thin lines to reveal unobstructed structures along lines of sight.}

We further assume each deblended velocity component is optically thin with respect to each other but not with respect to the CMB, individually. Our deblended model, with the CMB subtracted as a constant baseline, can thus be expressed as,

\begin{equation} \label{eq:deblend}
I_{\nu} = \sum_{j=i}^{m} \left[ B_{\nu}(T_{ex, j}) - B_{\nu}(T_{\mathrm{CMB}}) \right] \left[ 1 - e^{- \tau_{\nu, j}} \right],
\end{equation}
where $j$ designates each velocity component along a line of sight and $\tau_{\nu,j}$ is governed by,
\begin{equation} \label{eq:deblend_tau}
\tau_{\nu, j} =  0.1\ \tau_{0,j} \exp{\left[ \frac{-(v-v_{\mathrm{LSR}, j})^2}{2\sigma_{v,j}^2}\right]}.
\end{equation}
Here, $T_{ex,j}$, $\tau_{0,j}$, $v_{\mathrm{LSR}, j}$, and $\sigma_{v,j}$ are obtained from previously fitted models with hyperfine structures. To ensure the deblended emission have high spectral acuities for structure identification in ppv space, we further set $\sigma_{v,j}$ to 0.09 km s$^{-1}$ instead of adopting the line widths previously derived from our fits. This constant $\sigma_{v}$ value is roughly the minimal line width required to be Nyquist-sampled at our $0.07$ km s$^{-1}$ full-width-half-max (FWHM) spectral resolution. 

To illustrate structures revealed by deblending, Figure \ref{fig:ppvCubeAndSpine} shows the volume-rendered deblended ppv cube of our fits to the NH$_3$ (1,1) observations of NGC 1333.

\subsubsection{Identifying filaments}\label{subsub:fil_id}

We identify filaments in ppv space by using a multi-dimensional density ridge identification algorithm known as the Subspace Constrained Mean Shift (SCMS) \citep{Ozertem2011}. The mathematical framework behind SCMS was generalized by \cite{ChenYC2014arXiv} to operate on weighted, particle-like data in addition to their unweighted counterparts, which enables SCMS to run on gridded, multi-dimensional images. Since the publicly available \texttt{R} code developed by \cite{ChenYC2015MNRAS} for cosmological applications only implemented the original framework, we modified the code to reflect the generalized one. \edit1{We further translated the code to \texttt{Python}, parallelized it for multi-processing, and made it publicly available on \textit{GitHub} via the \textsc{CRISPy}\footnote{\texttt{CRISPy} codebase: \url{https://github.com/mcyc/crispy}.} (i.e., Computational Ridge Identification with SCMS for Python) package under a GNU General Public License. The version of \textsc{CRISPy} we used in this paper is archived in Zenodo \citep{CRISPyDOI}.}

While the \textsc{DisPerSE} algorithm (\citealt{Sousbie2011}; \citealt{Sousbie2011ptII}) recently used in star formation studies (e.g., \citealt{Arzoumanian2011}) also operates in 3D (e.g., \citealt{Smith2016}), it requires two more user-defined parameters to run than SCMS. The SCMS' ability to find density ridges consistently is also well established in statistical studies (\citealt{ChenYC2014arXiv}), making SCMS an attractive option over other methods. Furthermore, SCMS captures local information, such as ridge orientations, better than methods that derive ridges from monolithic filament objects (e.g., \citealt{Koch2015}) or approximate them as line segments (\citealt{Hacar2013}, \citeyear{Hacar2017}).

\begin{figure}
\centering
\includegraphics[width=0.41\textwidth,  trim={0 -3mm 0 0}]{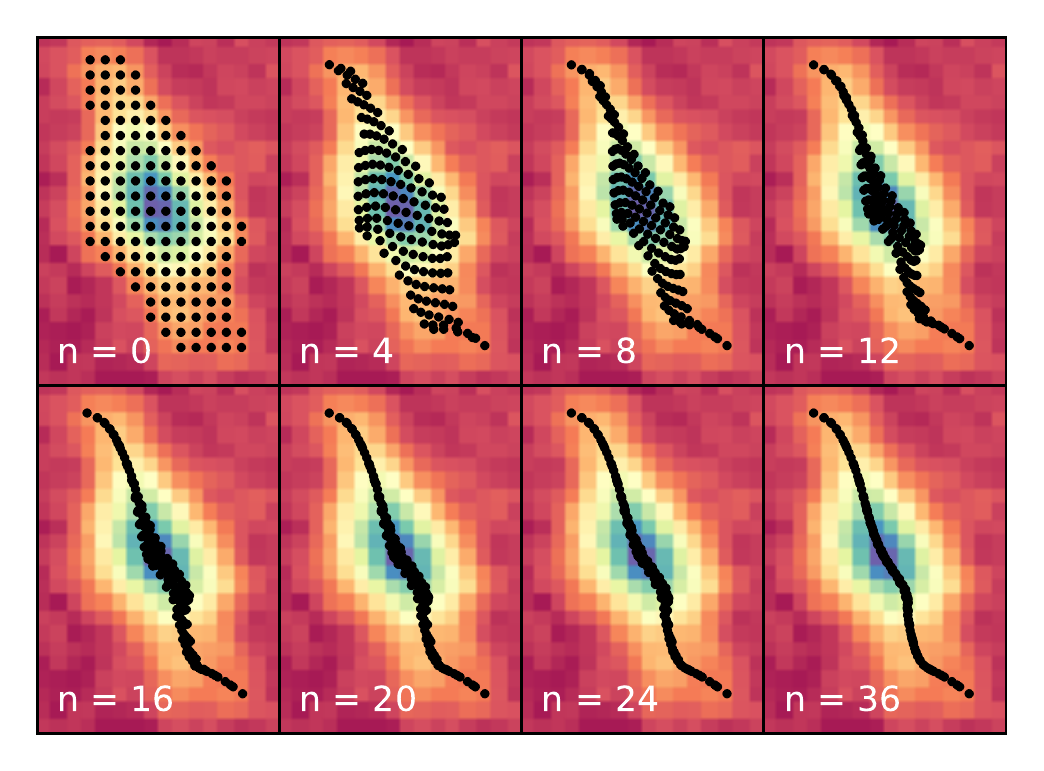}
\caption{Snapshots of SCMS finding a 2D `density' ridge from the image shown in the background, as carried out by \textsc{CRISPy}. The respective iteration number for each snapshot is labelled in each panel. The black dots represent the SCMS walkers and the background image was taken from the integrated NH$_3$ emission map of NGC 1333, cropped around a source in its north-east.}
  \label{fig:scms_steps}
\end{figure}

The SCMS algorithm defines a ridge as a smooth, continuous, one-dimensional object in a multi-dimensional density field. In addition to this nomenclature, we define a skeleton as a ridge that has been gridded onto an image and a spine as a skeleton with all its branches removed. The SCMS algorithm finds ridges by moving walkers iteratively up the density field using a gradient ascent method. This approach is subspace constrained (see \citealt{ChenYC2014arXiv}) to ensure the walkers converge on one-dimensional ridges instead of zero-dimensional peaks. Figure \ref{fig:scms_steps} demonstrates how SCMS identifies such a ridge in 2D from an NH$_3$ integrated intensity map of a source in NGC 1333 using the \textsc{CRISPy} package.

In general, the SCMS algorithm operates primarily on two user-defined parameters: density (e.g., intensity) threshold and smoothing bandwidth. The density threshold masks out noisy features in the density field while the smoothing bandwidth performs a kernel estimate of the field from particle-like data. Even though a gridded image, e.g., an emission cube, can in principle serve directly as a density field without a kernel estimate, a smoothing kernel is still required by the generalized SCMS to estimate density gradients efficiently and move its walkers accordingly \citep{ChenYC2014arXiv}. A smoothing length comparable to, or greater than, the resolution of the image is thus required still.

\begin{figure}
\centering
\includegraphics[width=0.44\textwidth,  trim={0 1mm 0 0}]{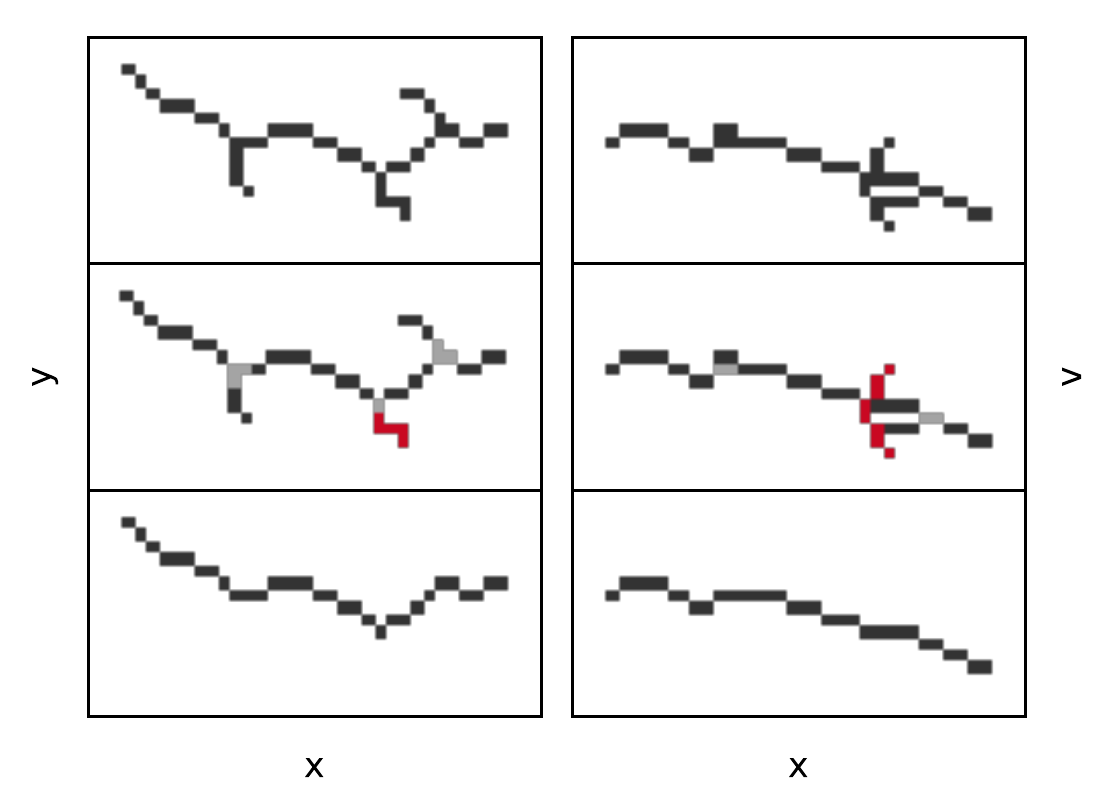}
\caption{A gridded, 3D skeleton being pruned into a spine, projected onto the xy and xv planes on the left and right panels, respectively. Top: the skeleton prior to pruning. Middle: the skeleton decomposed into branches (black), intersections (grey), and `bad' branches (red; see Section \ref{subsub:fil_id}). Bottom: the resulting spine, defined by the longest path in the skeleton with `bad' branches excluded.}
  \label{fig:prune_demo}
\end{figure}

For this work, we adopted a density threshold of 0.15 K and a smoothing length of 1.5 pixels for our SCMS run with \textsc{CRISPy}. The deblended cube was spatially convolved to twice its original beam width prior to the run. This additional step was performed to suppress noisy features further in the cube without sacrificing spectral resolution, particularly given that SCMS smooths its data indiscriminately in all three dimensions. Since SCMS operates natively in a continuous space, we set our convergence criterion such that the ridges identified are less than one voxel in width prior to re-gridding. We describe details on our choice of parameters further in Appendix \ref{apdx:scms}. 

Once \textsc{CRISPy} identified emission ridges in the continuous ppv space, we map these ridges back to the native grid of the deblended cube. These re-gridded ridges are referred to as skeletons and are subsequently pruned down to branchless structures we call spines. We accomplish such a pruning by using a graph-based technique developed by \cite{Koch2015}, which we have generalized to operate in 3D.

We prune branches by first decomposing a skeleton into intersection and branch objects known as nodes and edges in graph theory, respectively. We then find the longest path in the graph, measured in Euclidean distance, and subsequently remove all the edges outside of this path. To ensure our spines represent velocity-coherent structures, branches that may bridge velocity-discontinuities are further removed. We define these `bad' branches as ones with an on-sky length less than 9 pixels and a velocity projected length greater than that of its on-sky length in pixels (i.e., $\sim 4.8$ km s$^{-1}$ pc$^{-1}$ for NGC 1333 with our data). Figure \ref{fig:prune_demo} shows a demonstration of our pruning process with `bad' branches shown in red. We note that removing `bad' branches does not necessarily impose a maximum velocity gradient limit on a filament. This process merely breaks filaments apart.

\subsubsection{Assigning Membership to Filaments}\label{subsub:vSort}

We group our fits-derived velocity slabs obtained with \textsc{MUFASA} into velocity coherent structures by associating them with filament spines. In brief, we do so hierarchically by first placing velocity slabs into structures we call associations based on each slab's proximity to a spine in the ppv space. Such proximity is calculated using a spatial extension of a spine we call ppv-footprint, a structure for which the $v_{\mathrm{LSR}}$ separation between a slab and a spine can be referenced from at each pixel. This first step intends to disentangle filaments that overlap in projection into associations. 

Associations, which are allowed to have more than one velocity slab at each pixel, are then sorted internally to produce velocity coherent structures (vc-structures) that contain only a single slab at a pixel. This sortation is carried out based on kinematic similarities between the velocity slabs. The $v_{\mathrm{LSR}}$ map resulting from this sortation is subsequnetly median smoothed and adopted as the new ppv-footprint. This last step serves to grow and update the association iteratively starting from the filament spine, one pixel at a time.

\begin{figure}
\centering
\includegraphics[width=0.405\textwidth, trim={0mm 0mm 0 0}]{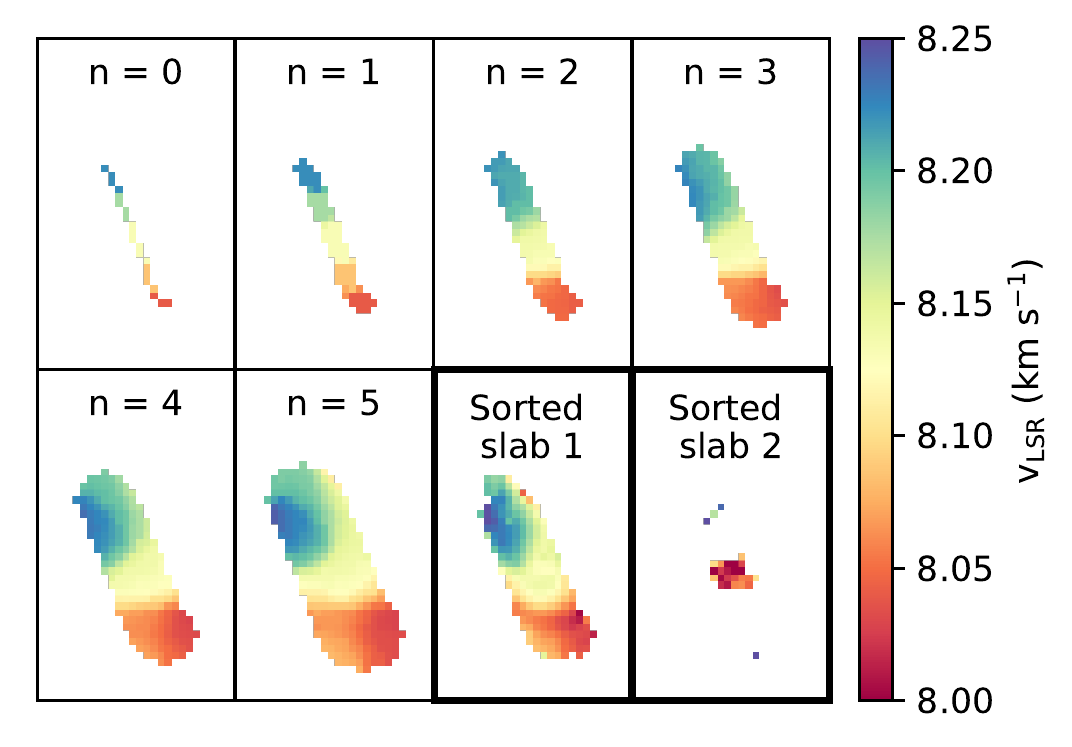}
\caption{A ppv-footprint during each iteration of its creation, shown in panels labelled with the iteration number \textit{n}. The v$_{\mathrm{LSR}}$ maps of the sorted velocity slabs, which have been assigned to the final ppv-footprint (n=5) as members of an association based on their $v_{\mathrm{LSR}}$ proximities, is shown in the last two panels, framed with bold borders. Only the first velocity slab represents a velocity coherent structure we used for our velocity gradient analyses.}
  \label{fig:ppv_footprint}
\end{figure}

We repeat such an assignation and sortation for five iterations. This number of iterations grows our vc-structures to an extent where SNR values of the new pixels start to drop off below 3. Figure \ref{fig:ppv_footprint} shows ppv-footprints corresponding to each of these iterations in panels labeled with the iteration number \textit{n}. The $v_{\mathrm{LSR}}$ maps of the first and second velocity slabs in the final association are shown in the last two panels of Figure \ref{fig:ppv_footprint}. This first velocity slab shown in the figure is representative of those that go into our final vc-structures, which are used in our velocity gradient analyses. Further details of our membership assignment to vc-structures are described in Appendix \ref{apdx:vcStructures}.

\subsection{Velocity Gradient Analysis}\label{subsect:VGradAnalysis}

\subsubsection{Decomposition of Vector Fields}\label{subsub:decompVecFields}

To study gas motions geometrically with respect to filament spines, we devised a technique to decompose a vector field (e.g., the velocity gradient field) into orthogonal components that are either parallel or perpendicular to a filament spine. Such a decomposition is accomplished by first taking a distance transform of a sky-projected spine to map out the shortest Euclidean distance between a given pixel and the spine. In other words, we calculated the radial distance between a pixel and a spine from which radial profiles of filaments can be constructed.

\begin{figure}
\centering
\includegraphics[width=0.45\textwidth, trim={7mm 5mm 0 2mm}]{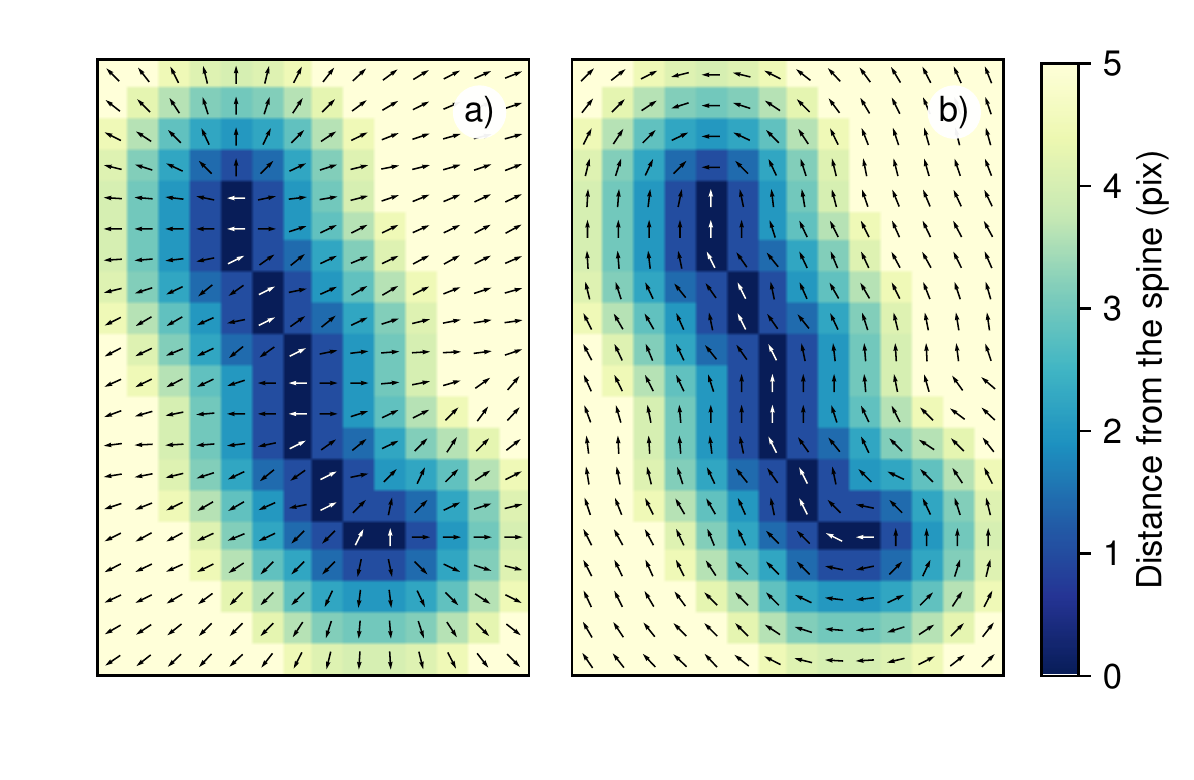}
\caption{Distance transform of a filament spine overlaid with its corresponding a) divergence field and b) parallel field. The vectors on and off the spine are colored white and black, respectively. The color scale corresponds to pixel distances between 0 and 5.}
  \label{fig:div_field_demo}
\end{figure}

Vector-fields that point orthogonally away from the filament spines are then created by taking gradients of our distance transforms over a sampling distance of 1 pixel using the second-order accurate central differences method. We refer to these fields as the divergence fields. Figure \ref{fig:div_field_demo}a shows an example of a divergence field superimposed on its corresponding distance transform.

Due to the sampling method of the gradient calculation and symmetry, the divergence field vectors right on the spine often have magnitudes of zero. To avoid a loss of information due to this limitation, we reconstructed an on-spine vector field parallel to the spine by taking gradients, i.e., central differences, of the spine's pixel coordinates. This on-spine field is then rotated by 90\textdegree \ and inserted into the divergence-field, as shown in Figure \ref{fig:div_field_demo}a in white.

\begin{figure*}
\centering
\includegraphics[width=0.37\textwidth]{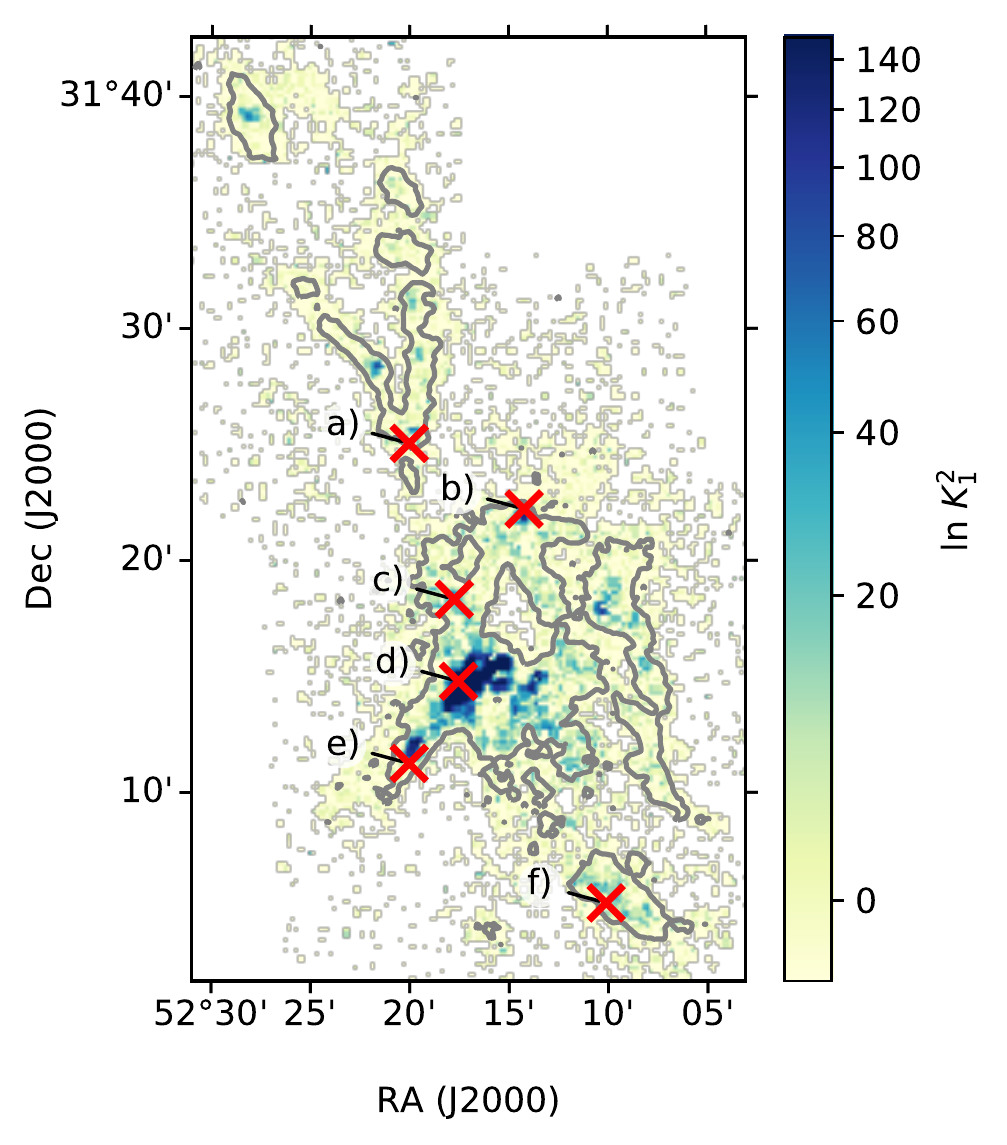}
\hspace{12mm}
\includegraphics[width=0.37\textwidth]{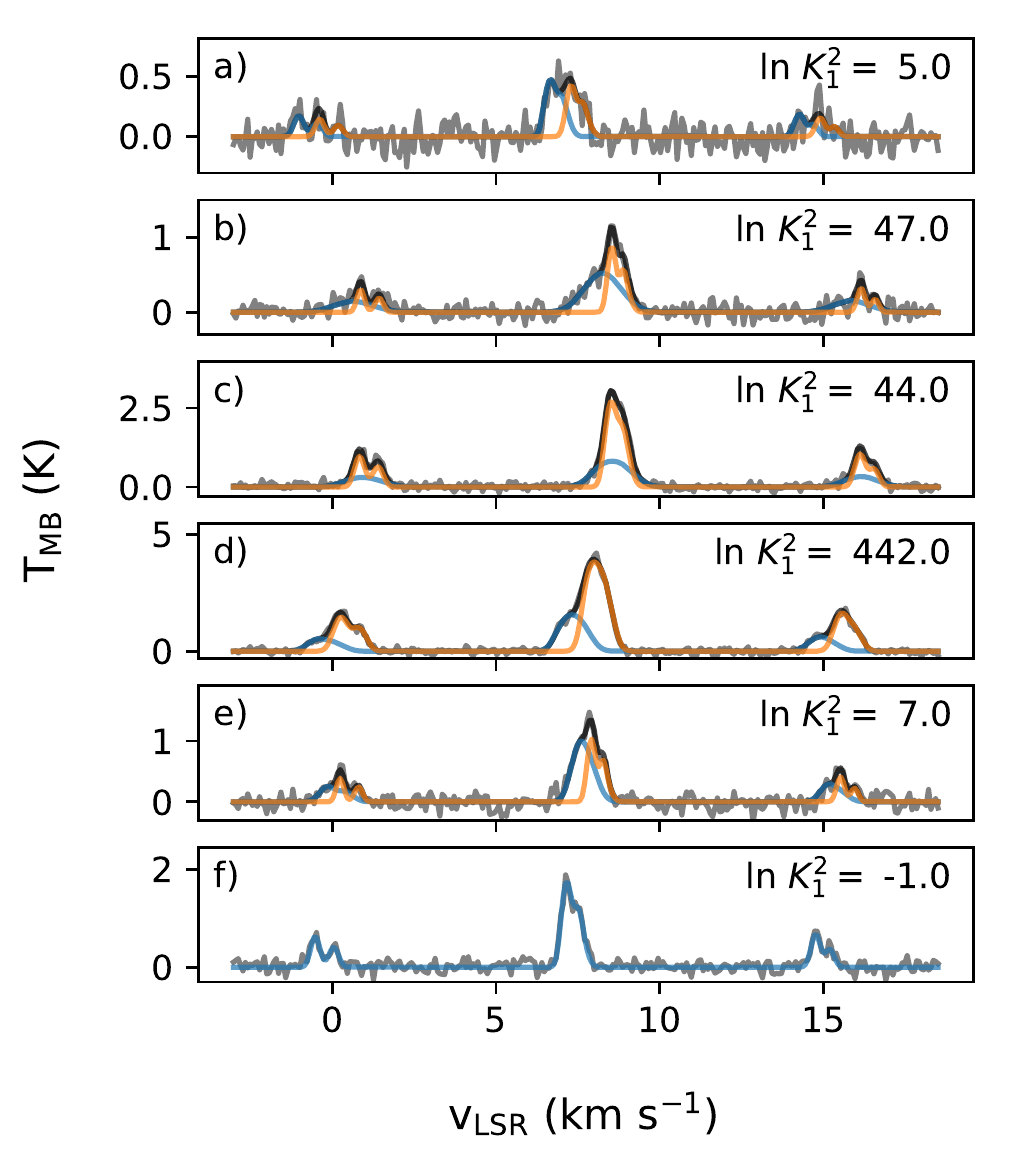}
\caption{Left panel: relative likelihood of the two-slab fit over its one-slab counterpart in NGC 1333. The grey contour shows the integrated NH$_3$ (1,1) emission at the 0.35 K km s$^{-1}$ level. Right panel: the observed NH$_3$ $(1,1)$ spectra (grey) extracted from the positions marked with red x's shown in the left panel, cropped to focus on the spectral regions near the main hyperfine lines. The spectra are superimposed with their corresponding two-slab fits (black) and models of their individual components (blue and orange).}
  \label{fig:likelihood_map}
\end{figure*}

We construct the corresponding parallel field of a filament by rotating the divergence field by 90\textdegree. Since the divergence field is discontinuous across the spine, where its vectors point oppositely away from each other, a uniform rotation of the divergence field will result in a discontinuous parallel field, where its vectors are anti-aligned with respect to each other across the spine. To address this issue, we rotated the divergence field vectors on each side of the spine independently in the directions opposite to each other. Such a rotation is accomplished by exploiting angular degeneracy in the $\arctan$ function to differentiate vectors on the two sides of the spine. Figure \ref{fig:div_field_demo}b shows an example of a parallel-field overlaid on the distance transform of its corresponding spine.

\subsubsection{Computing Velocity Gradients}\label{subsub:VGradCalc}

We calculate $v_{\mathrm{LSR}}$ gradients, i.e., $\nabla v_{\textup{LSR}}$, from $v_{\mathrm{LSR}}$ maps of velocity-coherent structures determined with the methods described in Section \ref{subsub:vSort}. These gradients are calculated on a pixel-by-pixel basis by fitting a plane over pixels within a 6-pixel diameter aperture centered on them. The diameter of the plane-fitting aperture is explicitly chosen to be twice the size of our FWHM beam to ensure the velocity gradients are calculated over resolved structures. To ensure the quality of our calculations, we calculate $\nabla v_{\textup{LSR}}$ only over apertures where $v_{\mathrm{LSR}}$ values are available for more than 1/3 of the pixels.

We further decompose the calculated $\nabla v_{\textup{LSR}}$ fields into components that are perpendicular and parallel with respect to its associated filament spine. This decomposition is accomplished by taking the dot products between the $\nabla v_{\textup{LSR}}$ field and the divergence field, as well as between the $\nabla v_{\textup{LSR}}$ field and the parallel field (see Section \ref{subsub:decompVecFields}).

\section{Results}\label{results}

Earlier in Section \ref{sub:linefit_tests}, we presented the performance of our fitting method, \textsc{MUFASA}, as characterized by our test fits to synthetic spectra. Here we present our best fit models to the GAS NH$_3$ (1,1) observations of NGC 1333 in Section \ref{sub:NGC1333_fit_results}, along with the deblended emission reconstructed from these fits. The filament spines identified from the deblended emission by \textsc{CRISPy} and the velocity slabs assigned to these spines are also presented in the same subsection. We further present our velocity gradient analysis on these velocity-coherent filaments in Section \ref{subsect:orientationVGrad}.

\subsection{NGC 1333 - Fitted Models}\label{sub:NGC1333_fit_results}

Figure \ref{fig:likelihood_map} shows the relative likelihood of the two-slab fit over the one-slab fit, i.e., $K^2_1$, in NGC 1333 as determined by the AICc (see Section \ref{subsect:multi-slab model}). A significant fraction, i.e., 40\%, of the pixels in NGC 1333 with SNR $> 3$ are determined to be better fitted with two-slab models based on the statistically robust threshold of $\ln{K^2_1} > 5$ \citep{Burnham2004}. This fraction is significantly higher than that suggested by the GAS DR1 paper \citep{Friesen2017}, where when examined by eye, only 5\% or less of the pixels with SNR $ > 3$ appears to be inadequately fitted by a one-component model. No pixel best fitted with our two-slab model has $\chi_{\nu} > 1.5$, which indicates that our observations of NGC 1333 are indeed well modelled with two or fewer velocity slabs.

\begin{figure*}
\centering
\includegraphics[width=0.304\textwidth]{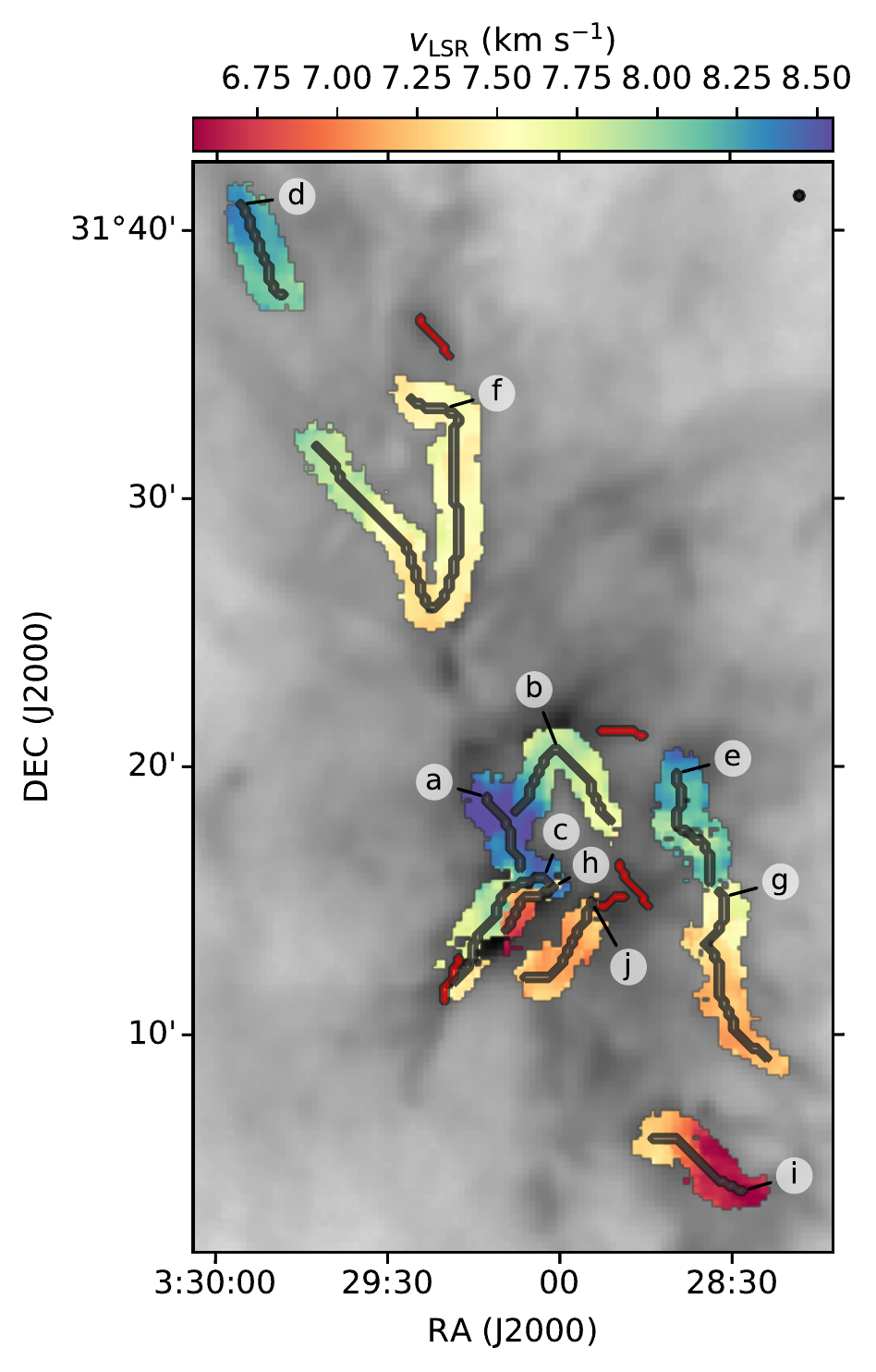} 
\includegraphics[width=0.28\textwidth]{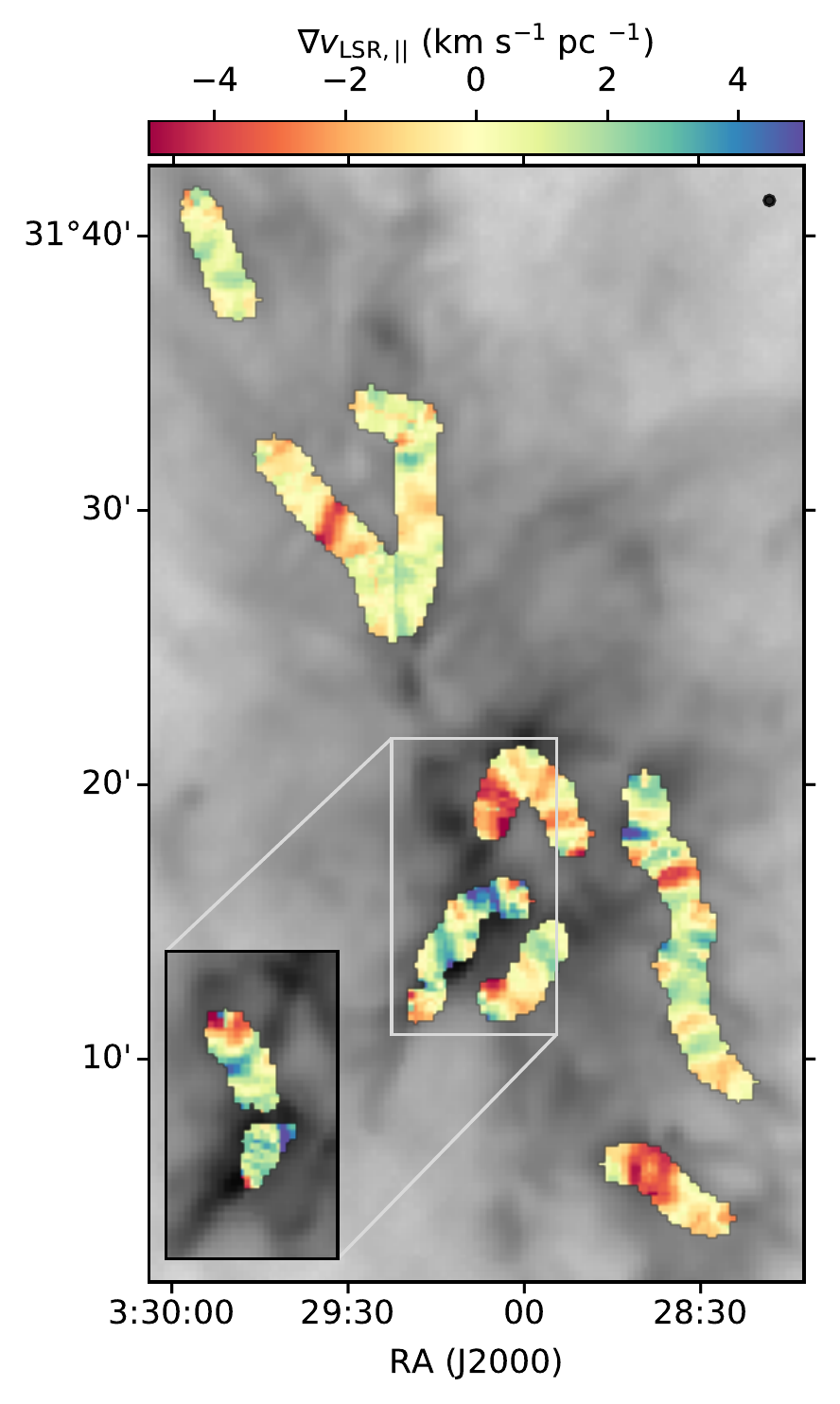}
\includegraphics[width=0.28\textwidth]{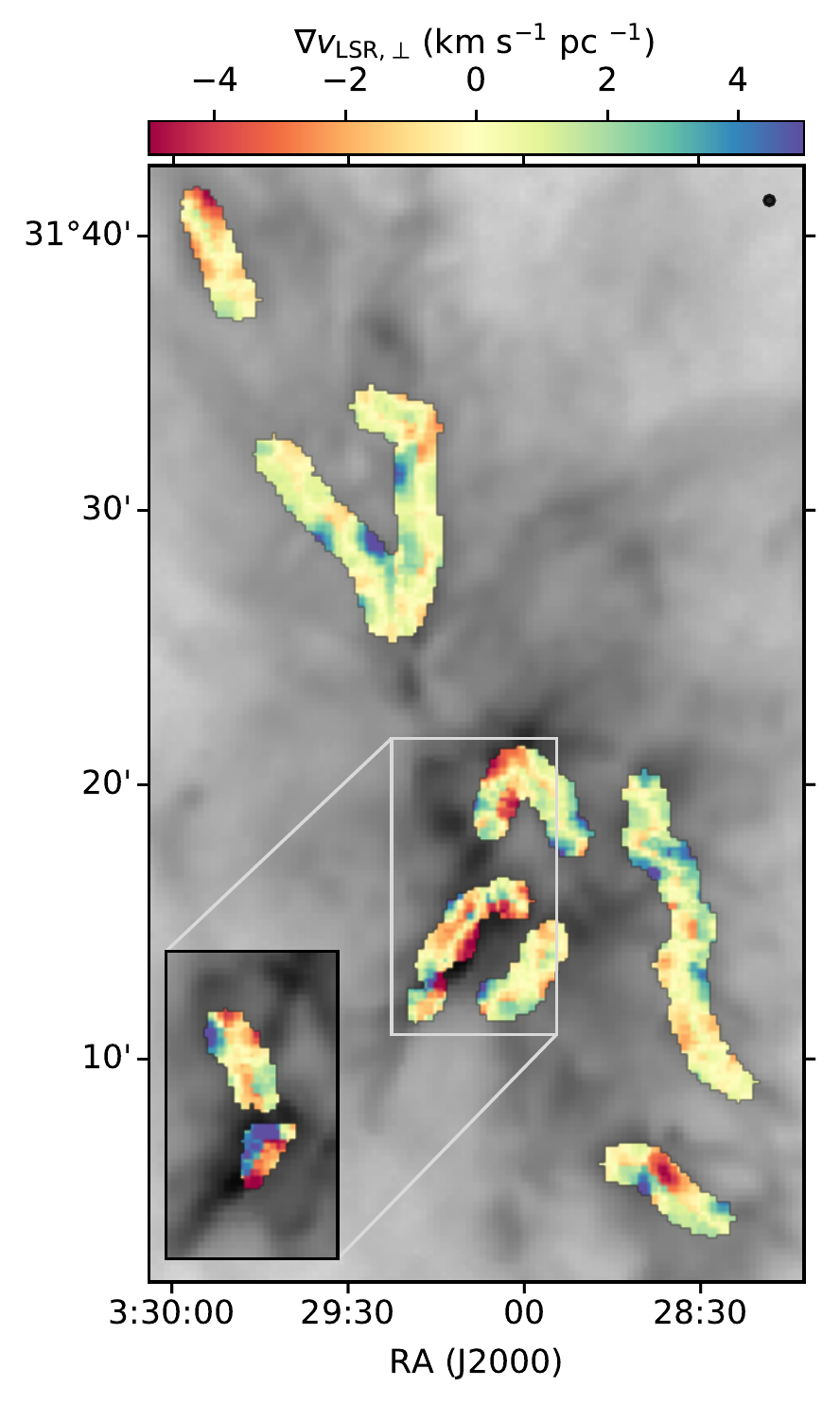}
\caption{Left panel: projections of filament spines identified in NGC 1333, overlaid on top of the $v_{\mathrm{LSR}}$ maps of selected velocity-coherent filaments (color). Spines less than 15 pixels in length are colored in red and their associated filaments are excluded from our analysis. Centre and right panels: spatial distribution of the parallel and perpendicular components of $\nabla v_{\textup{LSR}}$, respectively, relative to their filament spines. \edit3{The call-out boxes in these panels show the same $\nabla v_{\textup{LSR}}$ components of the additional, overlapping velocity-coherent filaments in the sky. The FWHM beam of the NH$_3$ (1,1) data (black circle) and the \textit{Herschel} N(H$_2$) map (grey; A. Singh et al., in prep.) are shown on the top right corner and the background of each panel, respectively.}}
  \label{fig:NGC1333_kin}
\end{figure*}

The right panel of Figure \ref{fig:likelihood_map} shows examples of our best fits to the NH$_3$ $(1,1)$ emission, superimposed on spectra extracted from positions marked in the left panel. It is qualitatively apparent that these spectra are indeed better fitted by two component models when $\ln{K^2_1} > 5$, even in the limiting case where $\ln{K^2_1}$ is near 5.

The deblended ppv cube of NGC 1333 derived from our best fit models is shown earlier in Figure \ref{fig:ppvCubeAndSpine}, overlaid with their respective spines identified by \textsc{CRISPy}. The left panel of Figure \ref{fig:NGC1333_kin} shows these spines projected onto the sky, and overlaid on top of the $v_{\mathrm{LSR}}$ maps of selected vc-structures, which are further overlaid on top of the \textit{Herschel} derived N(H$_2$) map (A. Singh et al., in prep). All the spines identified are used to sort fitted slabs into vc-structures, but only vc-structures with spines longer than 15 pixels in length ($\sim 6$ beam widths) are considered filaments in our analysis. \edit3{In NGC 1333, we identified 10 velocity-coherent filaments in total, and have labeled them alphabetically from \textit{``a''} to \textit{``j''} in Figure \ref{fig:NGC1333_kin}.}

\subsection{NGC 1333 - Velocity Gradients}\label{subsect:orientationVGrad}

The center and right panels of Figure \ref{fig:NGC1333_kin} show, respectively, the spatial distribution of perpendicular and parallel components of velocity gradients ($\nabla v_{\textup{LSR}}$) in the NGC 1333 filaments. These filaments display a wealth of $\nabla v_{\textup{LSR}}$ structures within them. A large fraction of these pixels have values of $\left | \nabla v_{\textup{LSR}} \right | > 2$ km s$^{-1}$ pc$^{-1}$ in both components, with many of them exceeding 4 km s$^{-1}$ pc$^{-1}$.

At our sampling distance of two-beam widths ($\sim 0.06$ pc in NGC 1333), our measured velocity gradients appear consistent with those median values reported by \cite{Hacar2013} in NGC 1333, measured with N$_2$H$^+$ in the parallel direction on the same spatial scale. Similarly, \cite{Lee2014} also reported comparable values in Serpens Main, measured in the parallel direction for filaments with mass $\sim 4$ M$_\odot$. Parallel gradients measured on larger scales ($> 0.2$ pc), however, tend to have smaller values. For example, the Serpens South filaments (\citealt{Kirk2013}; \citealt{Fernandez-Lopez2014}) and the Serpens Main filaments with masses of $\sim 15$ M$_\odot$ (\citealt{Lee2014}) all have larger-scale parallel gradients $ \leq$ 1.5 km s$^{-1}$ pc$^{-1}$.

\section{Discussion}\label{discussion}

\subsection{Comparing with N$_2$H$^+$ Analysis of NGC 1333}\label{subsect:2slabFitsNGC1333}

\cite{Hacar2017}, hereafter \citetalias{Hacar2017}, conducted a multi-component spectral analysis of NGC 1333 with a dense gas tracer, i.e., N$_2$H$^+$ (1-0) lines. Their data have a spatial and spectral resolution (30'' and 0.06 km s$^{-1}$, respectively) similar to our NH$_3$ (1,1) data, and a typical rms noise of 0.15 K, which is about 50\% higher than ours. \citetalias{Hacar2017} fitted their observations with a semi-automatic method, using either one- or two-component models as determined by eye.

About 15\% of the spectra successfully fitted by \citetalias{Hacar2017} are fitted with a two-component model. Considering not all of these successfully fitted spectra have SNR $> 3$, we estimate upwards to about 20\% of those spectra with SNR $> 3$ are fitted with two-component models. This estimate assumes all the successful two-component fits have SNR $>3$ in this limit. This 20\% fraction is significantly lower than that of our NH$_3$ (1,1) fits, where two-component models best fit about 40\% of our spectra with SNR $>3$. The difference in the model-selection criteria between our method (\textsc{MUFASA}) and that of \citetalias{Hacar2017} may contribute predominantly to this reported difference, where the conservative $\Delta v_{\mathrm{LSR}}$ threshold adopted by \citetalias{Hacar2017} may have culled out a significant fraction of their two-component fits.

\begin{figure}
\centering
\includegraphics[width=0.4\textwidth, trim={0 0 0 -12mm}]{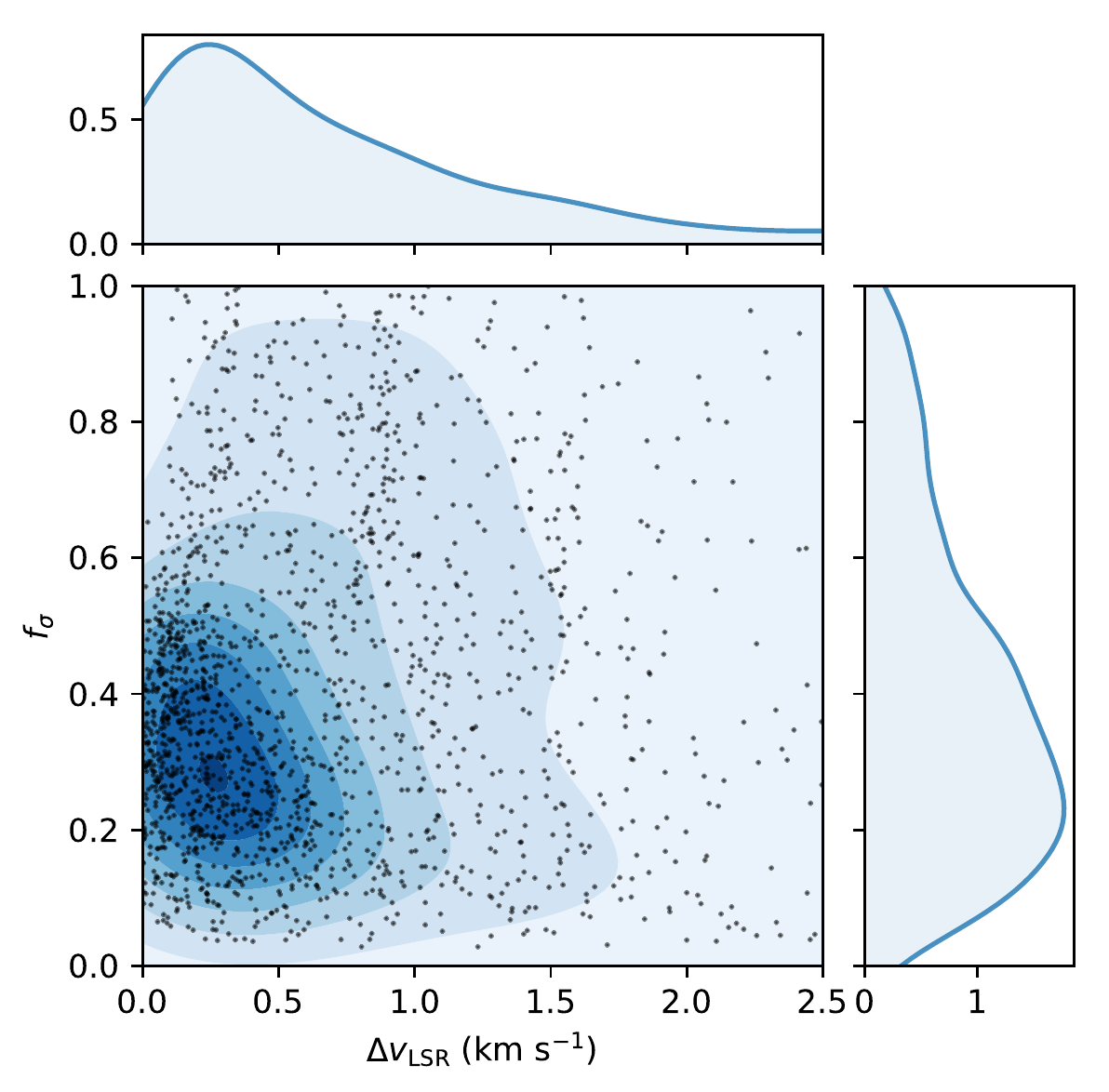}
\caption{The ratio between the fitted line widths of two velocity slabs (i.e., $f_{\sigma}$), plotted against velocity separation between these slabs ($\Delta v_{\mathrm{LSR}}$) for successful fits in NGC 1333. The colored background shows the contours of a kernel density estimate (KDE) while the filled curves to the top and the right show the 1D KDE distributions of $f_{\sigma}$ and $\Delta v_{\mathrm{LSR}}$, respectively.}
  \label{fig:fSig_vs_delV}
\end{figure}

Figure \ref{fig:fSig_vs_delV} shows the $f_{\sigma}$ and $\Delta v_{\textrm{LSR}}$ values derived from two-slab fits in NGC 1333 plotted against each other. The colored contours in the background show the kernel density estimate of this scatter plot. The filled curves to the right and the top of the plot show the 1D kernel estimated distributions of $f_{\sigma}$ and $\Delta v_{\mathrm{LSR}}$. Most of the points on this plot cluster around $f_{\sigma}$ and $\Delta v_{\mathrm{LSR}}$ values of 0.3 and 0.25 $\mathrm{km \ s}^{-1}$, respectively. This clustering places these values in the regime where the true-positive identification rate for two-slab spectra is in the range of $\sim 70 - 90$\% according to our truth test shown in Figure \ref{fig:2compID_vSep}b, accounting for all SNR values found in our synthetic test set. Given that \textsc{MUFASA}'s performance at identifying two-slab spectra decreases towards higher $f_{\sigma}$ values and lower $\Delta v_{\mathrm{LSR}}$ values, the true peak of the underlying two-slab population likely sits higher on the $f_{\sigma}$ axis and lower on the $\Delta v_{\mathrm{LSR}}$ axis. We reiterate that \textsc{MUFASA} only misidentifies a one-slab spectrum as a two-slab spectrum in $< 4\%$ of the test cases for all SNR values.

About 40\% of our two-slab fits to spectra with SNR $> 3$ in NGC 1333 have $\Delta v_{\textrm{LSR}}$ values that are less than 0.25 km s$^{-1}$, the threshold used by \citetalias{Hacar2017} to determine whether or not additional components are justified for their fits to N$_2$H$^+$ (1-0) observations of the same region. If NH$_3$ (1,1) and N$_2$H$^+$ (1-0) indeed trace the same gas population in this region, then the fraction of dense gas spectra in NGC 1333 with multiple velocity-components and SNR $>3$ may be significantly underreported by \citetalias{Hacar2017} due to their choice of $\Delta v_{\textrm{LSR}}$ threshold. Given that two-slab identification with \textsc{MUFASA} is \edit3{successful} even with moderate SNR values (i.e., 5 - 20; see Fig. \ref{fig:cmatrix}), the actual number of two-slab spectra with SNR $>3$ is likely higher than those reported in both the \citetalias{Hacar2017} study and our study here.

The NH$_3$ (1,1) and N$_2$H$^+$ (1-0) lines have critical densities of $\sim 2\times10^3$ cm$^{-3}$ and $\sim 5 \times 10^4$ cm$^{-3}$, respectively, at gas temperatures $\lesssim 20$ K (\citealt{Shirley2015}). The ratio between these critical densities remains similar even at higher temperatures. If the second velocity-component detected in our study tends to trace more diffuse gas, then the difference between the reported fraction of multi-component spectra between this work and that of \citetalias{Hacar2017} may be due to density differences in the tracers themselves in addition to the line-fitting methods used. The sensitivity difference between our data and that of \citetalias{Hacar2017} may also play a role as well. 

A recent high angular-resolution study of NGC 1333 concluded that NH$_3$ (1,1) and N$_2$H$^+$ (1-0) trace the same gas population well \citep{Dhabal2019}. Since this study only fits one velocity component along each line of sight, however, it is unclear how robust their conclusion is. Further investigation on how well NH$_3$ (1,1) and N$_2$H$^+$ (1-0) trace each other in NGC 1333 is thus needed, particularly for diffuse emission to which the data of \cite{Dhabal2019} are less sensitive. 

The filament spines we identified from our NH$_3$ data with \textsc{CRISPy} (see Figure \ref{fig:NGC1333_kin}, left) are morphologically similar to the `filament axes' identified by \citetalias{Hacar2017} with N$_2$H$^+$ observations. Some of the longer filaments, however, are `split' differently. Our filament \textit{f}, for example, is split into filaments \textit{12} and \textit{14} by \citetalias{Hacar2017}, while our filament \textit{g} is split into filaments \textit{1} and \textit{2} by \citetalias{Hacar2017}. Moreover, we identify a kinematically distinct filament (i.e., \textit{h}) that was not identified earlier by \citetalias{Hacar2017}, which runs closely parallel to our filament \textit{c} (i.e., \textit{10}).

Even though the spatial separation between spines of filament \textit{c} and \textit{h} is only slightly resolved in our data, the spectral separation of these spines ($\sim 0.9$ km s$^{-1}$) is well resolved. When observed at higher spatial resolutions with NH$_3$ and N$_2$H$^+$ \citep{Dhabal2019}, these two filaments can be seen by eye as distinct structures. Filament \textit{h} is thus likely missed by \citetalias{Hacar2017} due to their approach rather than observational biases introduced by the tracers used.

\subsection{Velocity Gradients on Large Scales}\label{subsect:vCohFilsNGC1333}

\begin{figure}
\centering
\includegraphics[width=0.47\textwidth, trim={0 0cm 0 0cm}]{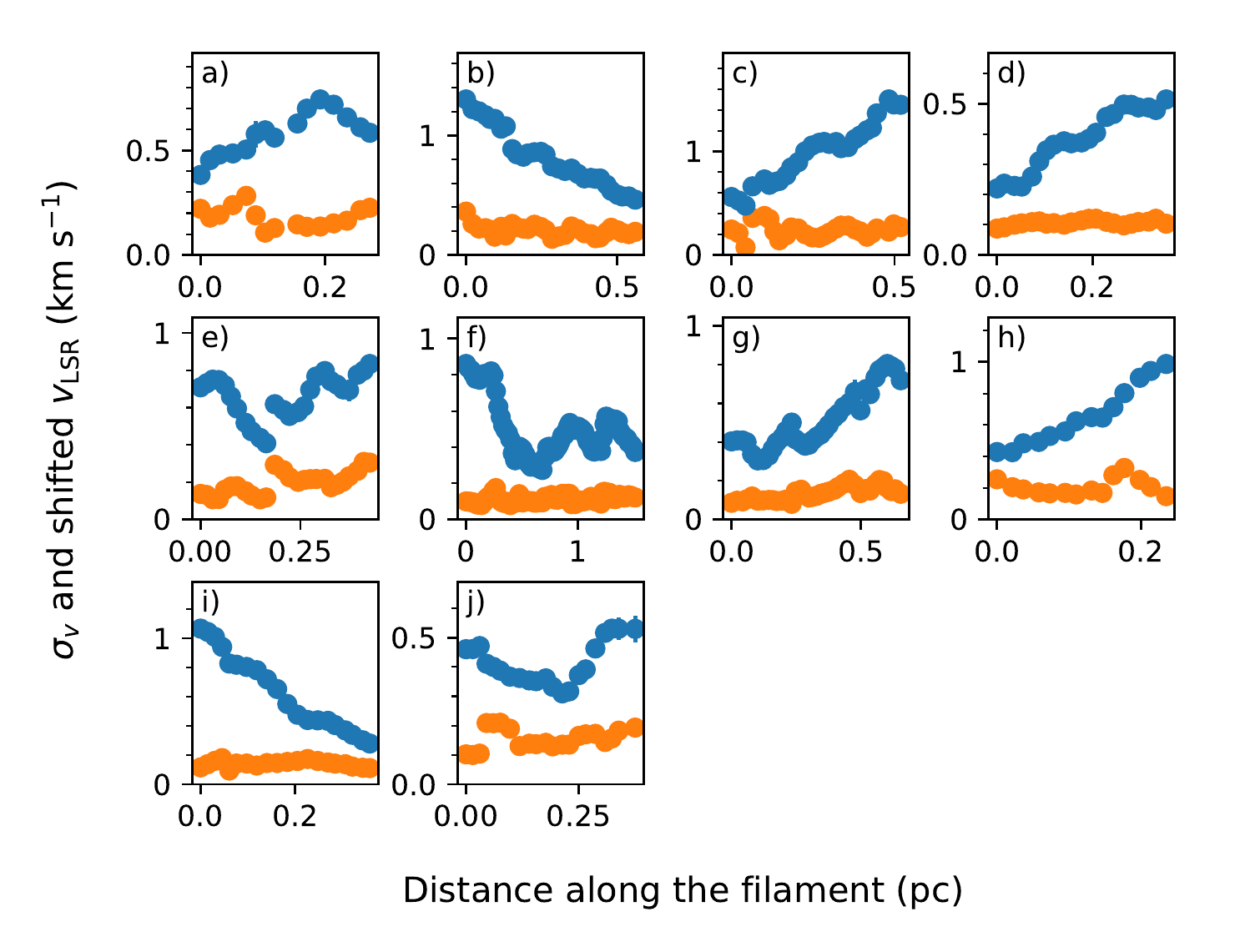}
\caption{\edit3{The profiles of $\sigma_{v}$ (orange dots) and $v_{\textup{LSR}}$ (blue dots) along filament spines identified in NGC 1333. The $v_{\textup{LSR}}$ shown here have been shifted to fit on the same axes as $\sigma_{v}$.}}
  \label{fig:len_pro_ngc1333}
\end{figure}

\edit3{Figure \ref{fig:len_pro_ngc1333} shows 1D $\sigma_{\mathrm{v}}$ and $v_{\mathrm{LSR}}$ profiles of the filaments identified in NGC 1333, taken directly from the pixels on their spines. The $v_{\mathrm{LSR}}$ values displayed in Figure \ref{fig:len_pro_ngc1333} have been the zero-point shifted arbitrarily from the local standard of rest (i.e., LSR) to fit nicely on the same axes as the $\sigma_{\mathrm{v}}$. We note that these spine profiles run in the direction that starts on the ends closest to the map origin, i.e., the south-eastern corner.}

The $v_{\mathrm{LSR}}$ variations along these spines are generally smooth with respect to their estimated errors. Only a few discontinuities are found in these profiles. Such a lack of discontinuity suggests that \textsc{CRISPy} does indeed identify density ridges robustly as continuous structures, which form the basis of our membership assignment to velocity-coherent filaments. 

The $v_{\mathrm{LSR}}$ values shown in Fig. \ref{fig:len_pro_ngc1333} do vary significantly along most of the identified filaments. Many of the $v_{\mathrm{LSR}}$ profiles are \edit3{approximately linear} and monotonic on scales larger than 0.2 pc, and have velocity gradients typically in the range of 0.8 - 2.5 km s$^{-1}$ pc$^{-1}$ on those scales. The large-scale ($> 0.2$ pc) velocity gradients are similar to those found with N$_2$H$^+$ observations in NGC 1333 ($\sim 0.5$ - 2.5 km s$^{-1}$ pc$^{-1}$, \citealt{Hacar2017}) and the Serpens South SF region (1.4 km s$^{-1}$ pc$^{-1}$, \citealt{Kirk2013}; $\sim 1$ km s$^{-1}$pc$^{-1}$, \citealt{Fernandez-Lopez2014}). These gradients also fall within the range measured in the Serpens Main SF region (0.7 - 4.8 km s$^{-1}$ pc$^{-1}$, \citealt{Lee2014}).

Velocity gradients along filaments on a large scale have often, but not uniquely, been interpreted as signatures of mass flow towards a star-forming core or cores (e.g., \citealt{Kirk2013}). While dense structures in NGC 1333 do not necessarily display a classical ``hub and spokes'' geometry, over-densities of dense cores and Class 0/I YSOs, as estimated by \cite{Hacar2017}, are often found at the ends of our filament spines in projection. Filaments \textit{c} and \textit{h}, for example, have very linear $v_{\mathrm{LSR}}$ profiles and large-scale $\nabla v_{\textup{LSR}}$ values of roughly 1.9 km s$^{-1}$ pc$^{-1}$ and 2.5 km s$^{-1}$ pc$^{-1}$, respectively. The fact that the north-western ends of these filaments coincide with the most prominent peak of over-densities in NGC 1333, in the SVS 13 vicinity (see \citealt{Walawender2008}), suggests these filaments are indeed transporting gas along their lengths towards a small (i.e., $n < 10$) cluster/proto-cluster. 

While filament \textit{b} has no end that correlates with an over-density of dense cores and Class 0/I YSOs, it does have an over-density midway through its length, located at the apex of its sharp turn near the HH 12 IR sources (i.e., VLA 42; see \citealt{Walawender2008}). Remarkably, this filament has a very linear and continuous $v_{\mathrm{LSR}}$ profile despite having such a distinct bend in its middle. Considering that this profile has a large-scale $\nabla v_{\textup{LSR}}$ of $1.4$ km s$^{-1}$ pc$^{-1}$, filament \textit{b} is likely a velocity-coherent system of two filaments that are transporting material towards a small hub.

Interestingly, the $\nabla v_{\textup{LSR}}$ seen on the largest scale of our observation ($\sim 4$ pc), i.e., at the clump scale, also appears to be fairly ordered along the north-south direction. Nearly all the filaments featuring linear $v_{\mathrm{LSR}}$ profiles along their spines have $v_{\mathrm{LSR}}$ values that increase northwards. Filament \textit{b} is the only exception, where half of its western segment prior to its sharp bend has $v_{\mathrm{LSR}}$ values that increase southwards instead. Even though filament \textit{f} does not have an overall linear $v_{\mathrm{LSR}}$ profile, its western portion prior to its sharp bend does have a segment, $\sim 0.2$ pc in length, with a fairly linear $v_{\mathrm{LSR}}$ profile and a $\nabla v_{\textup{LSR}}$ of $\sim 2.5$ km s$^{-1}$ pc$^{-1}$. The $v_{\mathrm{LSR}}$ values of this segment increase northwards as well.

In addition to the prevalent trend that $v_{\mathrm{LSR}}$ increases northward in most filaments, the median $v_{\mathrm{LSR}}$ value of each filament tends to increase northwards across the NGC 1333 clump as well. Considering that NGC 1333 is relatively elongated in the north-south direction on the clump scale ($> 4$ pc; see map by \citealt{Sadavoy2012} for example), most of these filaments may trace a larger filamentary inflow like those assumed by \cite{Matzner2015} in their model, of which these smaller filaments may be a part.

\subsection{Velocity Gradients on Small Scales}\label{subsect:vGrad_SmScales}

\subsubsection{Parallel Components}\label{subsub:vGrd_sm_parallel}

In addition to well-organized velocity structures on larger scales, Figures \ref{fig:NGC1333_kin} and \ref{fig:len_pro_ngc1333} reveal many quasi-oscillatory $v_{\mathrm{LSR}}$ structures that can be found on the 0.1 pc ($\sim 3$ beam widths) scale in NGC 1333. This behaviour is prominently visible along the spines of many filaments (see Figure \ref{fig:len_pro_ngc1333}) and shows up in the parallel $\nabla v_{\mathrm{LSR,} \parallel}$ map (see Figure \ref{fig:NGC1333_kin}, center) as ``zebra stripes.'' These small-scale structures, e.g., gradient peaks and dips, also appear to be somewhat evenly spaced by $\sim 0.1$ pc, which suggests a quasi-oscillatory wavelength of $\sim 0.2$ pc. Interestingly, this behaviour is not confined to the spines of filaments and extends spatially across the width of a filament.

Similar quasi-oscillatory $v_{\mathrm{LSR}}$ behaviours have been found in Taurus L1495/B213 (\citealt{Tafalla2015}) and in Serpens South filaments (\citealt{Fernandez-Lopez2014}) based on one-component fits to N$_2$H$^+$ (1-0) observations. Filaments and fibres identified from multi-component fits to C$^{18}$O (1-0) observations in Taurus-Auriga L1517 (\citealt{Hacar2011}), and Taurus L1495/B213 (\citealt{Hacar2013}), respectively, also showed similar results. These quasi-oscillatory $v_{\mathrm{LSR}}$ behaviours generally resemble those seen in synthetic C$^{18}$O observations of simulated filaments, constructed with various degrees of realism (e.g., \citealt{Moeckel2015}; \citealt{Smith2016}; \citealt{Clarke2018}).  

We find no strong spatial correlations between quasi-oscillatory $v_{\mathrm{LSR}}$ and dense structures in NGC 1333. This result is contrary to that found in Taurus-Auriga L1517 (\citealt{Hacar2011}) using C$^{18}$O observations but agrees with the behaviour found in Taurus L1495/B213 (\citealt{Tafalla2015}) using N$_2$H$^+$ (1-0) observations. This agreement extends to synthetic C$^{18}$O observations of simulations (e.g., \citealt{Smith2016}). \edit3{The lack of correlation between quasi-oscillatory $v_{\mathrm{LSR}}$ values and dense structures suggests the former is not driven by periodic gravitational instabilities.} \edit3{Alternative mechanisms, such as magnetic waves explored by \cite{Tritsis2016}, \cite{Tritsis2018}, and \cite{Offner2018}, may be responsible for these quasi-oscillatory behaviours.}

\subsubsection{Perpendicular Components}\label{subsub:vGrd_sm_parallel}

Filaments in NGC 1333 also contain a wealth of perpendicular velocity gradients, i.e., $\nabla v_{\textup{LSR,}\perp}$, structures on smaller scales (see right panel of Fig. \ref{fig:NGC1333_kin}). Regions with high $| \nabla v_{\textup{LSR,}\perp} |$ values ($> 2$ km s$^{-1}$ pc$^{-1}$) tend to form spatially compact but resolved $\nabla v_{\textup{LSR,}\perp}$ structures on the outskirts of the filaments, i.e., away from the spine. \edit1{Similar to interpretations made in the literature (e.g., \citealt{Palmeirim2013}; \citealt{Dhabal2018}),} these compact $\nabla v_{\textup{LSR,}\perp}$ structures may be indicative of recent or ongoing accretion of nearby gas onto the filaments.

Free-fall accretion in analytic models typically has estimated infall velocities of a few km s$^{-1}$ at filament `boundaries' (e.g., \citealt{Palmeirim2013}; \citealt{Heitsch2013}). Such an infall velocity will likely result in shocks if the accreting filament is in hydrostatic equilibrium like those described in classic models (e.g., \citealt{Stodolkiewicz1963}; \citealt{Ostriker1964}). Even in numerical models for which filaments are not equilibrium substructures, shock-induced discontinuities in velocities are expected from accretion (e.g., \citealt{Clarke2018}).

We neither saw nor expected velocity discontinuities in our filaments because velocity-coherent structures are continuous in their velocities by definition. Velocity discontinuities, however, can be inferred from the $\Delta v_{\mathrm{LSR}}$ observed between velocity slabs along a line of sight. With the exception of filaments \textit{c} and \textit{h}, which have typical $\Delta v_{\mathrm{LSR}}$ values of $\sim 0.8$ km s$^{-1}$ between them, we did not find filaments with overlapping velocity slabs that had $\Delta v_{\mathrm{LSR}}$ values greater than 0.4 km s$^{-1}$, i.e., about twice the isothermal sound speed at 10 K. Interestingly, the region where filaments \textit{c} and \textit{h} overlap along lines of sight is also where some of the most prominent $\nabla v_{\textup{LSR,}\perp}$ structures are found in NGC 1333. We note that this observed $\nabla v_{\textup{LSR,}\perp}$ structure is unlikely driven by the highly collimated outflow originated from IRAS 4 (e.g., \citealt{Blake1995}) given that it poorly aligns with the orientation of the outflow.

Except for where filaments overlap in projection, we typically only detect two-slab spectra near filament spines rather than the edges. This lack of detection near the edges is likely limited by the sensitivity of our data. Not much information is therefore available on $\Delta v_{\mathrm{LSR}}$ over filament edges to infer the nature of spatially compact $\nabla v_{\textup{LSR,}\perp}$ structures that reside there. Furthermore, despite having a critical density of $~10^{3}$ cm$^{-3}$, \edit3{it remains unclear how effective NH$_3$ is at tracing accretion flows, which themselves are likely more diffuse than the dense filaments.} Further investigation with NH$_3$ synthetic observations, similar to that conducted by \cite{Clarke2018} with C$^{18}$O transitions, would be highly valuable.

To search for potential sources of accretion flows, we looked for structures around our filaments in the column density, i.e., N(H$_2$), map of NGC 1333 derived from \textit{Herschel} observations (A. Singh et al., in prep). We find no strong spatial correlation, however, between ambient \textit{Herschel} structures (e.g., sub-filaments) and the observed $\nabla v_{\textup{LSR,}\perp}$ structures. This lack of correlation suggests that accretion from sub-filaments, such as those seen in Taurus B211/B213 (\citealt{Goldsmith2008}, \citealt{Palmeirim2013}), is unlikely to explain the origin of the compact $\nabla v_{\textup{LSR,}\perp}$ structures seen near filament edges.

Nevertheless, the lack of visible, interconnected ambient structures does not necessarily rule out $\nabla v_{\textup{LSR,}\perp}$ as a sign of accretion flows onto dense filaments in NGC 1333. According to models where a post-shock layer of a converging flow produces filaments (e.g., \citealt{ChenOstriker2014}; \citeyear{ChenOstriker2015}), a sub-filamentary network only arises in a strong magnetic field. In these models, gravity drives the accretion flows in a post-shock layer. The resulting flows move predominately along the field lines and may not necessarily contain sub-structures with densities high enough to be distinguished from the rest of the planar, accretion flow. Without visible substructures, these flows may appear as a large-scale background to \textit{Herschel} due to their planar geometry, making them difficult to discern. 

Post-shock accretion models, such as those developed by \cite{ChenOstriker2014}, have been proposed by \cite{Dhabal2019} as an explanation for the observed large $\nabla v_{\textup{LSR,}\perp}$ along the south-western edge of filament \textit{c}. This particular $\nabla v_{\textup{LSR,}\perp}$ structure has been found in both the high resolution NH$_3$ and N$_2$H$^+$ observations by \cite{Dhabal2019} as well as our NH$_3$ observations. The filament \textit{h} we identify with two-slab fits, which runs parallel to filament \textit{c}, also displays similar $\nabla v_{\textup{LSR,}\perp}$ over the same region. In a post-shock accretion interpretation, such a similarity suggests that filament \textit{h} belongs to the same planar flow as filament \textit{c}. Interestingly, filament \textit{h} is spatially well resolved in the high-resolution observation by \cite{Dhabal2019} as a filament distinct from \textit{c}.

It is worth noting that \cite{Walsh2006} measured infall velocities of $\sim 1$ km s$^{-1}$ towards the south-western edge of filament \textit{h} with HCO$^+$ (1-0) observations, modelled as self-absorbed lines. The infall velocities measured with HCO$^+$ (1-0), which were interpreted as a sign of large-scale ($\sim 0.2$ pc) infall, are similar to the observed $v_{\mathrm{LSR}}$ separation ($\sim 0.8$ km s$^{-1}$) between filaments \textit{c} and \textit{h} in NH$_3$ along lines of sight. Given that this infall region spatially correlates with filaments \textit{c} and \textit{h}, the HCO$^+$ (1-0) observed there may indeed trace the same planar accretion flow as that suggested by the large NH$_3$ $\nabla v_{\textup{LSR,}\perp}$ we see towards filaments \textit{c} and \textit{h}.

Not all the observed $\nabla v_{\textup{LSR,}\perp}$ structures in our filaments can be well explained by models of accretion flow along a post-shock layer. While the compact nature of $\nabla v_{\textup{LSR,}\perp}$ may be explained by clumpy, inhomogeneous accretion, the sign (i.e., direction) alternation of $\nabla v_{\textup{LSR,}\perp}$ along filament edges, however, does not conform well to the planar geometry naively expected from a post-shock layer. Some, if not all, of these observed $\nabla v_{\textup{LSR,}\perp}$ features, may thus be driven by a different physical process. 

\edit1{While rotation of small bodies, such as dense cores, may produce compact $\nabla v_{\textup{LSR,}\perp}$ signatures, no clear correlation exists between cores and many of these compact $\nabla v_{\textup{LSR,}\perp}$ regions in the majority of the cases. Inhomogeneous accretion flows, on the other hand,} similar to those seen in the non-magnetized simulation by \cite{Clarke2017} may explain the sign-changing behavior of these $\nabla v_{\textup{LSR,}\perp}$ along filament edges. Indeed, the shocked regions bordering dense structures in their simulation, as traced by local velocity divergence, morphologically resemble the $\nabla v_{\textup{LSR,}\perp}$ structures seen in NGC 1333. Further investigation with synthetic observations of simulations is needed to see if such a resemblance holds and whether or not magnetic fields play an important role in such an accretion. 

\subsubsection{Radial Dependencies}\label{subsub:vGrad_rad_dependency}

\begin{figure}
\centering
\includegraphics[width=0.48\textwidth, trim={0 0cm 0 0cm}]{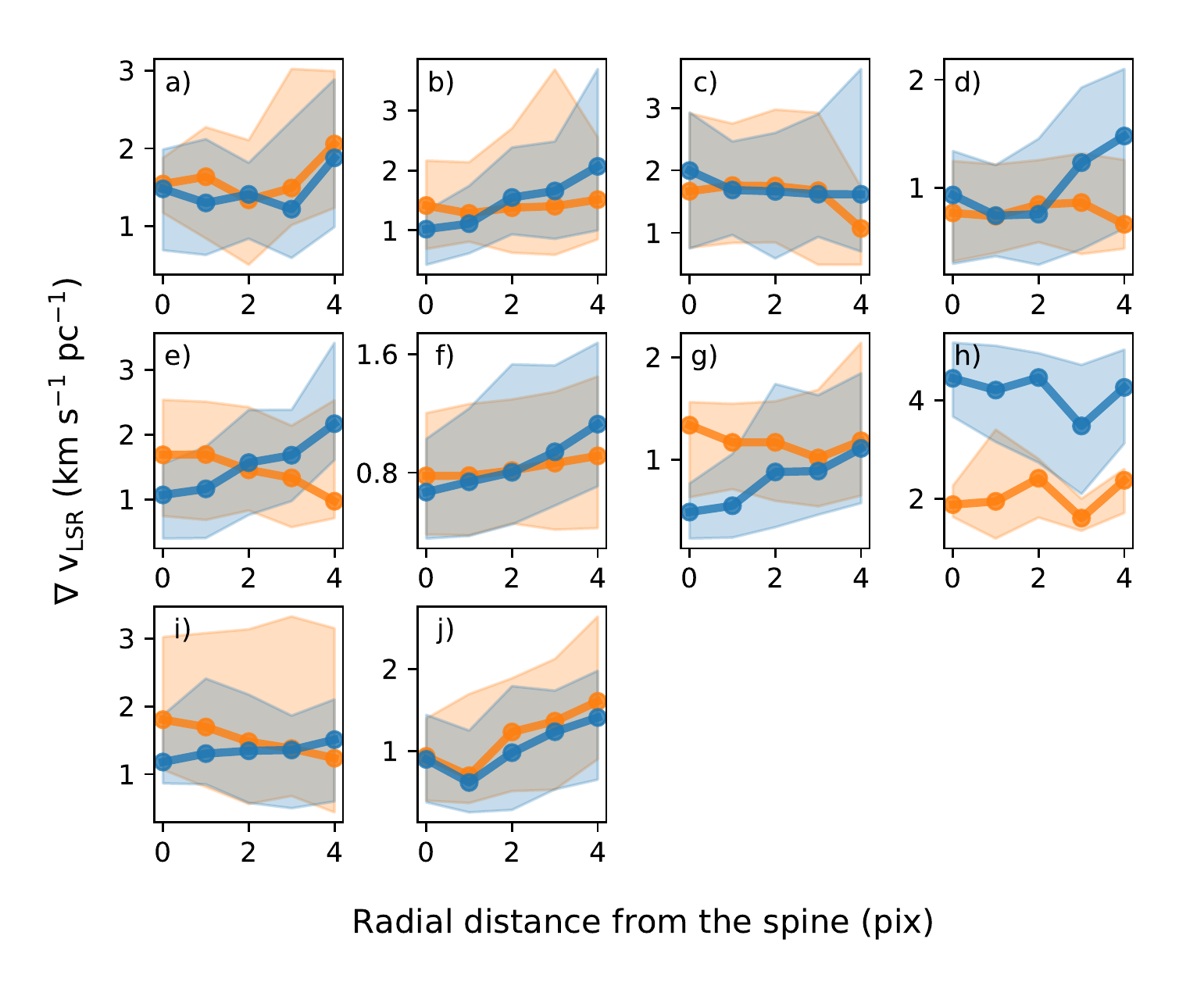}
\caption{Median magnitudes of the perpendicular (blue) and parallel (orange) velocity gradients plotted as a function of their distance to the filament spines. The median values are marked by the connected dots while the \nth{15} and \nth{85} percentile values mark the lower and upper bound of the shaded regions, respectively.}
  \label{fig:radial_vgradMag_ngc1333}
\end{figure}

Figure \ref{fig:radial_vgradMag_ngc1333} shows the magnitudes of the perpendicular and parallel velocity gradients plotted as functions of their respective distances from the filament spines. Their median, \nth{15}, and \nth{85} percentile values are marked. While the parallel velocity gradients show little correlation with their distances from filament spines, the perpendicular velocity gradients tend to decrease as one moves towards the spine in many filaments. 

\edit3{Specifically, filaments \textit{b}, \textit{e}, \textit{f}, \textit{g}, and \textit{j} clearly show such a trend. Indeed, the Wald Test (see \citealt{Fahrmeir2013}) revealed that these filaments all have p-values $< 0.01$ for a null hypothesis where the data are consistent with a zero-slope linear trend. The linear least-squares regression slopes of these trends are 0.40, 0.41, 0.17, 0.21, and 0.16, respectively, in units of km s\textsuperscript{-1} pc \textsuperscript{-2}. Moreover, the Pearson correlation coefficients, i.e. r-values (see \citealt{Cohen1988}), of these regressions all fall in the range of 0.2 - 0.4, suggesting that these positive correlations are indeed significant but relatively scattered.}

The decreasing trends in these filaments persist when gradients are calculated with smaller sampling distances ($r < 3$ pixels). Indeed, the regression slopes for these filaments actually increases with smaller sampling distances.

A decrease in $|\nabla v_{\textup{LSR,}\perp}|$ towards a filament spine contradicts the behaviour of free-falling gas. Such a free-fall behavior is often assumed for gas accretion onto a filament in simple analytical models (e.g., \citealt{Heitsch2009}; \citealt{Palmeirim2013}). While filaments themselves are not typically expected to behave like pressureless systems, except for maybe the very massive ones (e.g., M $\sim 600$ M\textsubscript{\(\odot\)}; \citealt{Gomez2014}), \edit3{a simple analytical examination of such an assumption can still provide vaulable perspectives in light of our results.}

Consider a parcel of gas falling onto an infinitely long cylinder axially centered at $r=0$. Such a parcel would have a velocity profile of $v_{\mathrm{ff}} \propto  [ \ln(r_{0}/r) ]^{1/2}$ if the parcel was initially at rest at $r_{0}$ (see \citealt{Heitsch2009}). The radial derivative of this profile, i.e., $\nabla v_{\textup{LSR,}\perp}$, thus would be $dv_{\mathrm{ff}}/dr \propto [r^2 \ln(r_{0}/r) ]^{-1/2}$ and would increase monotonically in magnitude towards the filament spine for $r$ values less than $\sim r_{0}/2$. The observed $\nabla v_{\textup{LSR,}\perp}$ for that parcel of gas should thus increase towards a filament spine \edit3{if the parcel's emission dominates over others along the line of sight, and its trajectory is not parallel to the plane of the sky.}

\edit1{While such a geometric assumption does not describe a simple, axially-symmetric accretion of a filament, it reasonably approximates inhomogeneous (e.g., \citealt{Clarke2017}) or anisotropic (e.g., \citealt{ChenOstriker2014}) accretion flows found in realistic simulations. In fact, the former model is unlikely to produce observable $\nabla v_{\textup{LSR,}\perp}$ in the first place, contrary to what we have observed.} Thus, if these non-symmetric assumptions hold true for our observations, then the observed decrease in $|\nabla v_{\textup{LSR,}\perp}|$ towards the filament spine indeed suggests these filaments do not behave like a pressureless system \edit1{under the accretion flow interpretation}. The $\nabla v_{\textup{LSR,}\perp}$ we observed may thus suggest ongoing accretion that is being damped by the higher density material as the accreting gas moves closer to the filament spine.

\subsubsection{Orientations}\label{subsub:vGrad_orientations}

\begin{figure}
\centering
\includegraphics[width=0.48\textwidth, trim={0 0.5cm 0 0cm}]{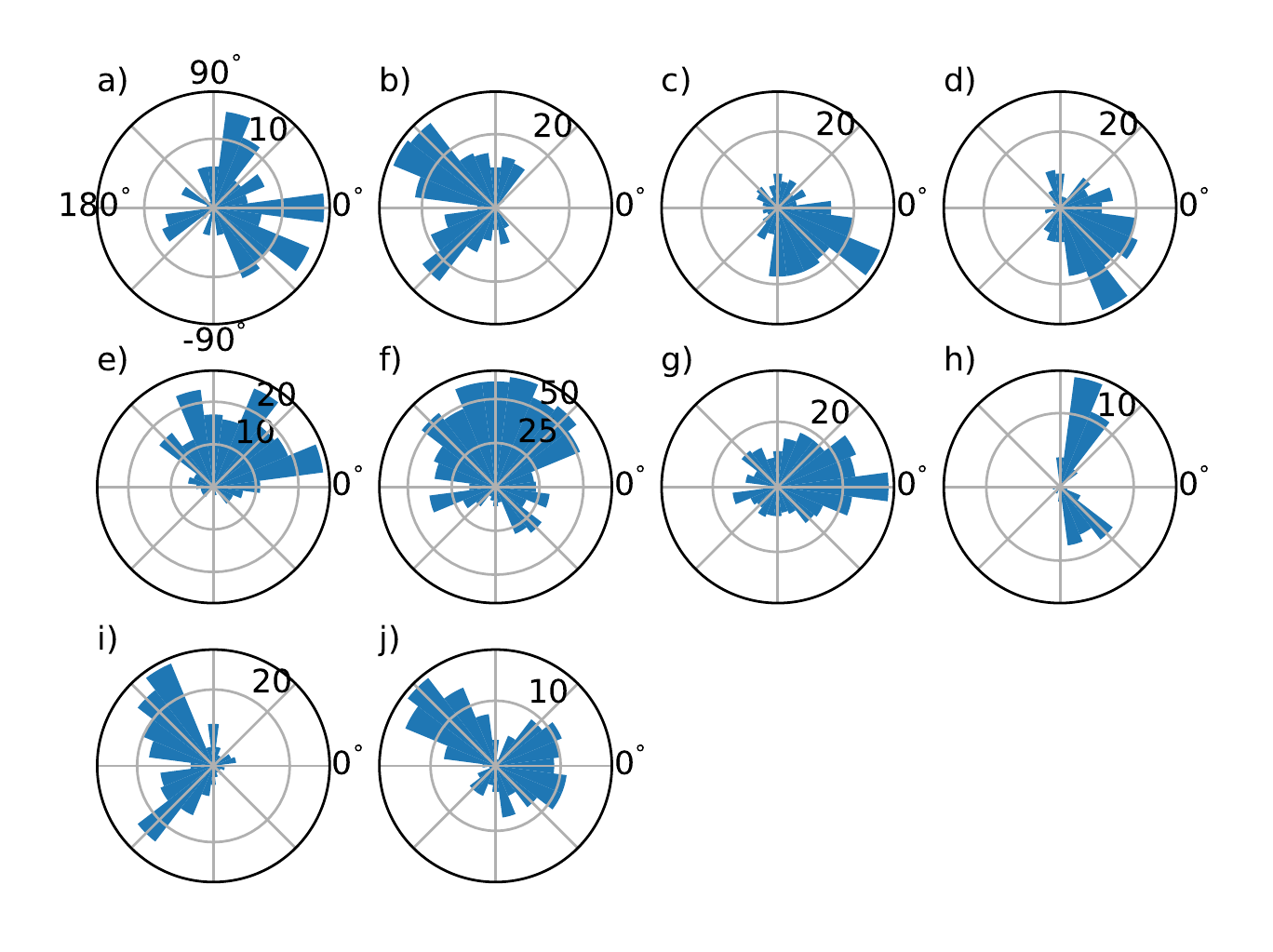}
\caption{Polar histograms of the $\nabla v_{\textup{LSR}} $ orientations shown for each filament.} 
  \label{fig:polar_hist_ngc1333}
\end{figure}

Figure \ref{fig:polar_hist_ngc1333} shows the orientation angles of the measured velocity gradients, $\theta$, binned into polar histograms. The direction along a filament spine away from the end closest to the image origin, i.e., the bottom left corner, defines the zero-point reference of our angle $\theta$. The convention is set such that vectors with $-180^{\circ}< \theta < 0^{\circ}$ point away from the spine and vectors with $0^{\circ}< \theta < 180^{\circ}$ point towards the spine. 

Most of the $\nabla v_{\textup{LSR}}$ vectors within an NGC 1333 filament are not randomly oriented and often display unimodal or bimodal behaviours. A circular statistics analysis conducted with a Rayleigh test (\citealt{Wilkie1983}) shows that the $\theta$ values found in 9 out of the 10 of filaments are very unlikely to have been drawn from a random distribution ($p < 0.01$). 

Coherent $\nabla v_{\textup{LSR}}$ orientations may seem perplexing at first considering how complex the $\nabla v_{\textup{LSR}}$ structures appear on small scales. A clearer picture emerges, however, when the $\nabla v_{\textup{LSR}, \parallel}$ on large scales and the radial dependency of $|\nabla v_{\textup{LSR,}\perp}|$ on small scales are considered together. For example, mass flows along filaments combined with perpendicular (i.e., radial) accretion onto filament edges may have caused the preferential $\nabla v_{\textup{LSR}}$ orientations observed here. After all, physics that are likely important in forming filaments, e.g., gravity and magnetic fields, do tend to impose order. If the observed $\nabla v_{\textup{LSR}}$ values are indeed indicative of mass flows, then filaments in NGC 1333 may be viewed as loci of collapsing flows where radially accreted gas changes direction to flow along filaments and into cores.

\section{Summary}\label{summary}

In this paper, we devised an efficient and robust method (i.e., \textsc{MUFASA}) to fit two-slab NH$_3$ (1,1) spectra automatically, one that is generalizable for other molecular species. We tested the performance of \textsc{MUFASA} on synthetic spectra and found it to be particularly robust at identifying two-component spectra with dissimilar $\sigma_v$ between their components. This behavior makes \textsc{MUFASA} \edit3{effective} at disentangling subsonic gas from supersonic gas along lines of sight. By selecting our best fit model via a statistical approach, we recovered $\sim 40\%$ of the two-component NH$_3$ spectra from our synthetic data with $\Delta v_{\mathrm{LSR}} $ values below the 0.25 km s$^{-1}$ culling threshold used by \cite{Hacar2017} for their study of NGC 1333 with N$_2$H$^+$. Based on our fit residuals, we find no strong evidence that three-component fits are warranted in NGC 1333. 

We identified 3D filament spines in ppv space from \textsc{MUFASA} line-fitted models using our implementation of the generalized (i.e., weighted) SCMS algorithm (see \citealt{ChenYC2014arXiv}), \textsc{CRISPy}. The generalized SCMS operates on a well-established statistical framework where the orientations of the density ridges (e.g., spines) are well defined locally. We sorted the fitted models into velocity-coherent filaments with these spines and measured the velocity gradients of their filaments on a beam-resolved scale ($\sim 0.05$ pc). We further decomposed these velocity gradients into parallel and perpendicular components with respect to the local spine.  

By applying our analysis techniques to the observation of NGC 1333, we find the following:
\begin{enumerate}  

\item Many filaments have remarkably linear changes in $v_{\mathrm{LSR}}$ along their spines on larger scales ($>0.2$ pc). The $\nabla v_{\mathrm{LSR}}$ corresponding to these changes are in the range of 0.8 - 2.5 km s$^{-1}$ pc $^{-1}$, similar to those found in previous works (e.g., \citealt{Kirk2013}). Several of these filaments have ends or sharp bends that spatially correlate with the kernel estimated over-densities of dense cores and embedded YSOs derived by \cite{Hacar2017}. This result suggests the observed $\nabla v_{\mathrm{LSR}}$ may indeed be tracing mass flow along filaments towards star-forming cores.

\item Most filaments with a linear $v_{\mathrm{LSR}}$ profile along their spines have values that increase northwards. Considering that the median $v_{\mathrm{LSR}}$ of these filaments tends to increase northwards as well, these filaments may trace a larger-scale ($> 4$ pc) filamentary accretion flow, similar to that assumed by \cite{Matzner2015} in their model, from which the NGC 1333 clump may be fed.

\item The $\nabla v_{\mathrm{LSR,}\parallel}$ measured on small scales show quasi-oscillatory $v_{\mathrm{LSR}}$ along filaments. These quasi-oscillations, however, do not correlate well with dense structures, a result similar to those found by \cite{Tafalla2015} in real N$_2$H$^{+}$ observation and by \cite{Smith2016} in synthetic C$^{18}$O observations of a simulation. \edit3{This lack of correlation suggests periodic gravitational instabilities are not responsible for such an observed behaviour. Alternative mechanisms, such as magnetic waves (e.g., \citealt{Tritsis2016}) may be responsible instead.} 
 
\item The $\nabla v_{\mathrm{LSR,}\perp}$ found on small scales tend to form compact structures near the filament edges, potentially indicating perpendicular accretion flows. The compact nature of these $\nabla v_{\mathrm{LSR,}\perp}$ structures combined with an apparent lack of ambient sub-filaments suggest these accretion flows are likely clumpy, i.e., inhomogeneous. Alternations in the direction of these $\nabla v_{\mathrm{LSR,}\perp}$ structures along filament edges also suggest these accretion flows may not be purely planar like those found in simulated magnetized post-shock layers (e.g., \citealt{ChenOstriker2014}), except possibly those first reported by \cite{Dhabal2019} for filaments \textit{c} and \textit{h}.

\item The magnitudes of the measured $\nabla v_{\mathrm{LSR,}\perp}$ decrease prominently toward filament spines in half (i.e., 5) of our filaments. \edit1{Assuming our observations trace gas flows that are inhomogeneous or anisotropic with respect to the filament spines, such a trend is inconsistent with free-fall accretion models and suggest that these filaments do not behave like pressureless structures.} Such an observed behaviour may thus indicate the infall of accretion flows being damped by the denser, pressure-supported gas within filaments.

\item The $\nabla v_{\mathrm{LSR}}$ vectors measured on small scales are not randomly oriented within a filament. Their orientations tend to be unimodally or bimodally distributed. This global trend within filaments conforms to a scenario by which the gas falling onto a filament is redirected to flow along the filament as it approaches the spine. 

\end{enumerate}
For our interpretations, we assumed the observed velocity gradients are indeed signs of accelerating gas seen along lines of sight. We plan to apply our analysis to synthetic NH$_3$ observations in future work to understand better the nature of these velocity gradients.


\acknowledgments

MCC, JDF, EWR, and JK acknowledge the financial support of a Discovery Grant from NSERC of Canada. The Green Bank Observatory is a facility of the National Science Foundation operated under cooperative agreement by Associated Universities, Inc. The National Radio Astronomy Observatory is a facility of the National Science Foundation operated under cooperative agreement by Associated Universities, Inc. JEP acknowledge the financial support of the European Research Council (ERC; project PALs 320620). AP acknowledges the financial support of the the Russian Science Foundation project 19-72-00064. MCC would like to further thank the support of Ben and Jerry's ice cream, especially the `Tonight's Dough,' that got him through long nights of writing.

%

\vspace{5mm}
\facilities{GBT (KFPA)}

          
\software{APLpy \citep{APLpy2012}, Astropy \citep{Astropy2018}, CRISPy \citep{CRISPyDOI}, FilFinder \citep{Koch2015}, Matplotlib \citep{Matplotlib2007}, MUFASA \citep{MUFASADOI}, NumPy \citep{NumPy2011}, scikit-image \citep{vanderWalt2014}, SciPy \citep{SciPy2001}, PySpecKit \citep{Ginsburg2011}, Python} 



\appendix\label{appendix}
\section{Parameter choices for \textsc{CRISPy}}\label{apdx:scms}

As mentioned in Section \ref{subsub:fil_id}, the operation of SCMS, including our implementation in \textsc{CRISPy}, primarily depends on two user-defined parameters: density (i.e., intensity) threshold and smoothing bandwidth. We adopted a density threshold of 0.15 K to capture most of our emission in the model while avoiding going near the typical rms noise level of our data ($\sim 0.1$ K). We further adopted a smoothing length of 1.5 pixels, which corresponds to about the 1-$\sigma$ sampling width of our data, i.e., $\sim 3$ pixels across the FWHM beam. Since our deblended cube at its native resolution is too noisy for SCMS even after a density cutoff, we further smoothed our deblended cube spatially to twice its original beam width prior to running SCMS. We smoothed the data only spatially and not spectrally to avoid further loss in our spectral resolution.

In addition to density threshold and smoothing bandwidth parameters, SCMS requires a few additional parameters to run in practice: convergence criterion, the maximum number of iterations, and walker placement. Convergence criterion and the maximum number of iterations are used to decide when to stop running SCMS further. We set our convergence criterion to $10^{-3}$ to ensure the ridges represented by the converged walkers have scatters that are smaller than the equivalent width of a voxel in the deblended cube. We set the maximum number of iterations to 1000 and the walker placement such that a walker is placed at each voxel in the deblended cube above a density threshold. We adopted a walker placement threshold of 0.16 K to sample the density field well without placing walkers near the edges of the field defined by our cutoff threshold (0.15 K).

The \textsc{CRISPy} implementation of SCMS also includes scaling parameters for which each dimension of the deblended emission can be rescaled. The purpose of such a rescaling is to renormalize the distance between each particle in the field and consequently the density field. Such a renormalization is essential for structure identification in a parameter space with dimensions that are not necessarily physically related, e.g., a ppv space.

Since the two spatial distances in a ppv space are physically related, only the velocity axis needs to be rescaled, provided that the smoothing bandwidth was already chosen appropriately based on the spatial sampling. For our runs, we kept velocity scaling the same as the one native to our deblended cube. We made this choice deliberately to avoid elongating spatially compact structures along the velocity axis such that they are misidentified as filaments. Shorter scaling is avoided to prevent further loss in our spectral resolution from bandwidth smoothing.


\section{Membership Assignment to Velocity Coherent Structures}\label{apdx:vcStructures}

As briefly described in Section \ref{subsub:vSort}, velocity slabs are assigned memberships to velocity-coherent structures (vc-structures) based on their proximity to filament spines in the ppv space. This process is performed iteratively, starting with pixels spatially closest to the filament spine. At each iteration, velocity slabs are assigned to spine-derived ppv-footprints nearest to them in $v_{\mathrm{LSR}}$ along a line of sight, which forms what we call an association. The slabs in each association are subsequently sorted internally, based on their kinematic similarities, into a vc-structure that contains only a single velocity slab along a line of sight. Here, we describe the process of assigning membership to vc-structures in more detail.

We first create a ppv-footprint, a spatial extension of a spine, to serve as a reference from which the $v_{\mathrm{LSR}}$ proximities between a slab and a spine are calculated along a line of sight, i.e., a pixel. We construct a ppv-footprint by dilating, i.e., expanding, filament spines by one pixel in the two spatial dimensions but not in the velocity dimension. This expansion is initially accomplished by taking the first-moment map of a filament spine, dilating the map's on-sky footprint by one pixel, and adopting the median $v_{\mathrm{LSR}}$ value of the moment map within a 3-pixel radius of each pixel as its new value.

Once the ppv-footprint is created, velocity slabs with $v_{\mathrm{LSR}}$ values closest to the ppv-footprint at each pixel are then assigned to that ppv-footprint as a member of the association. Only slabs with velocity separations $< 0.21$ km s$^{-1}$ from the ppv-footprint, i.e., about three spectral channel widths, are accepted to ensure the assigned members are reasonably velocity-coherent with respect to the ppv-footprint. We note that this threshold implicitly imposes an upper limit to velocity gradients of 0.42 km s$^{-1}$ pix$^{-1}$ (i.e., $\sim 29$ km s$^{-1}$ pc$^{-1}$) for a given association.  

Following these assignments, member slabs within each association are further sorted into a vc-structure containing only a single slab along each line of sight, based on their similarities in $v_{\mathrm{LSR}}$, $\sigma_{v}$, and $\delta v_{\mathrm{LSR}}$, i.e., the Jacobian estimated error of $v_{\mathrm{LSR}}$ from the fits. We used $\delta v_{\mathrm{LSR}}$ similarity as our additional proxy for spatial coherence assuming that spectral components which are spatially similar in their properties, such as their brightness, will produce fitted $\delta v_{\mathrm{LSR}}$ that are spatially similar as well. 

At each iteration, we sort these velocity slabs as follows:
\begin{enumerate}  
\item Assign velocity slabs with the smallest $\delta v_{\mathrm{LSR}}$ values at each pixel in a given association into a new, single slab structure we call a vc-structure.
\item Median-smooth the $\delta v_{\mathrm{LSR}}$ map of the vc-structure with a circular aperture, 1 pixel in radius, to serve as a reference map.
\item Reassign velocity slabs at each pixel with the most similar $\delta v_{\mathrm{LSR}}$ values, i.e., the least difference between the smoothed $\delta v_{\mathrm{LSR}}$ map and their respective $\delta v_{\mathrm{LSR}}$ values, to the vc-structure.
\item Create reference maps of $v_{\mathrm{LSR}}$ and $\sigma_{v}$ by employing the same median-smoothing technique used in step 2 on property maps of the vc-structure,
\item Reassign velocity slabs with the minimum quadrature sum difference between the reference maps and their respective $v_{\mathrm{LSR}}$ and $\sigma_{v}$ values, i.e., $\left[ (v_{\mathrm{LSR}} - v_{\mathrm{LSR, ref}})^2 + (\sigma_{v} - \sigma_{v, \mathrm{ref}})^2 \right]^{0.5}$, to the vc-structure.
\end{enumerate}

Following each sortation above, we median smooth the final $v_{\mathrm{LSR}}$ map in each vc-structure with a 3 pixel radius aperture and adopt the resulting $v_{\mathrm{LSR}}$ map as our new ppv-footprint for the subsequent iteration of membership assignment and sortation. This procedure is carried out for five iterations in total, growing ppv-footprints and their respective associations radially one pixel at a time. This five pixel radial distance typically marks the spatial extent for which the SNR values of our pixels start to fall below 3.

To illustrates such a process, Figure \ref{fig:ppv_footprint} shows a ppv-footprint at each iteration in panels labeled with the iteration number \textit{n}. The last two panels of Figure \ref{fig:ppv_footprint} show $v_{\mathrm{LSR}}$ maps of the first and second velocity slabs in an association, taken from the final iteration. The first slab shown here defines the final vc-structure used in our filament analyses. 


\bibliography{bibliography}{}
\bibliographystyle{aasjournal}



\end{document}